# An Efficient and Distribution-Free Two-Sample Test Based on Energy Statistics and Random Projections

Cheng Huang and Xiaoming Huo

**Abstract**

A common disadvantage in existing distribution-free two-sample testing approaches is that the computational complexity could be high. Specifically, if the sample size is $N$, the computational complexity of those two-sample tests is at least $O(N^2)$. In this paper, we develop an efficient algorithm with complexity $O(N \log N)$ for computing energy statistics in *univariate* cases. For *multivariate* cases, we introduce a two-sample test based on energy statistics and random projections, which enjoys the $O(KN \log N)$ computational complexity, where $K$ is the number of random projections. We name our method for *multivariate* cases as Randomly Projected Energy Statistics (RPES). We can show RPES achieves nearly the same test power with energy statistics both theoretically and empirically. Numerical experiments also demonstrate the efficiency of the proposed method over the competitors.

## 1 Introduction

Testing the equality of distributions is one of the most fundamental problems in statistics. Formally, let $F$ and $G$ denote two distribution function in $\mathbb{R}^p$. Given independent and identically distributed samples

$$\{X_1, \dots, X_n\} \text{ and } \{Y_1, \dots, Y_m\}$$

from two unknown distribution $F$ and $G$, respectively, the two-sample testing problem is to test hypotheses

$$\mathcal{H}_0 : F = G \text{ v.s. } \mathcal{H}_1 : F \neq G.$$

There are a few recent advances in two sample testing that attract attentions in statistics and machine learning communities. [17] propose a test statistic based on the optimal non-bipartite matching, and, [3] develop a test based on shortest Hamiltonian path, both of which are distribution-free. [8] develop a kernel method based on maximum mean discrepancy. [21], [22] and [1] consider a test statistic based on pairwise distance within the same sample and across two different samples, which also motivates this work.

Computational complexity is a common limitation in the aforementioned methods. Let $N = n + m$ denote the size of the two-sample testing problem. The Cross Match (CM) test in [17] requires solving the non-bipartite matching problem, whose computational complexity is: (1) $O(N^3)$ with optimal solution, see [7]; (2) $O(N^2)$ with greedy heuristic. The two-sample test in [3] is based on shortest Hamilton path, which is an NP-complete problem, and its computational complexity is $O(N^2 \log N)$ with heuristic method based on Kruskal's algorithm ([13]). The Maximum Mean Discrepancy (MMD) proposed by [8] requires computing the kernel function values of all pairs of samples, whose complexity is $O(N^2)$. Similarly, the energy statistics based methods in [21] and [1] typical requires the pairwise Euclidean distance, which also costs $O(N^2)$ complexity.

As a summary, the computational complexity of the aforementioned two-sample tests is at least $O(N^2)$, which leads to substantial computing time and prohibits their feasibility when the sample size $N$ is too large. As a solution, we develop an efficient algorithm for computing the energy statistics in [21] with complexity $O(N \log N)$ for univariate random variables. For multivariate random variables, we propose an efficient algorithm of complexity $O(KN \log N)$ with the technique of random projection, where $K$ is the number of random projections. The main idea of the multivariate algorithm is as follows: firstly, we project the data along some random direction; then, we use the univariate fast algorithm to compute the energy statistics with the univariate projected data; lastly, we repeat previous procedure for multiple times and take the average. As we will show in Theorem 4.12, the proposed test statistic based on random projections has nearly the same power with energy statistics.

The technique of random projection has been widely used in two-sample testing problems. [15] propose a new method, which firstly projects data along a few random direction; and then, applies the classical Hotelling $T^2$ statistic, for testing the equality of means in different samples. [20] develop a similar approach based on random projection and the Hotelling $T^2$ statistic, but the random projection is taken with respect to sample mean vectors and sample covariance matrices. These two papers focus on the problem under multivariate Gaussian settings while our work is more general and does not impose any assumptions in the distributions.

The remainder of this paper is organized as follows. We will review the definition and property of energy distance and energy statistics in Section 2. In Section 3, we describe the details of fast algorithms and corresponding two-sample tests. Asymptotic properties of proposed test statistic will be studied in Section 4. In Section 5, we will some numerical examples with simulated data to illustrate the computational and statistical efficiency of the proposed test. Discussions could be found in Section 6 and we will conclude in Section 7.

Throughout this paper, we adopt the following notations. We denote $c_p = \frac{\pi^{(p+1)/2}}{\Gamma((p+1)/2)}$ and $C_p = \frac{c_1 c_{p-1}}{c_p} = \frac{\sqrt{\pi}\Gamma((p+1)/2)}{\Gamma(p/2)}$ as two constants, where $\Gamma(\cdot)$ denotes the Gamma function. We also denote $|\cdot|$ as the Euclidian norm. For any vector $v$, $v^T$ is its transpose.

## 2 Review of Energy Distance and Energy Statistics

Energy distance is initially proposed by [21] to measure the distance between two multivariate distributions. We follow the definition of energy distance in [23].

**Definition 2.1.** *[23, Definition 1] Suppose $X, Y \in \mathbb{R}^p$ are two real-valued independent random variables with finite means, i.e., $\mathbb{E}[|X|] < \infty$ and $\mathbb{E}[|Y|] < \infty$, then the energy distance between $X$ and $Y$ is defined as*

$$\mathcal{E}(X,Y) = 2\mathbb{E}[|X-Y|] - \mathbb{E}[|X-X'|] - \mathbb{E}[|Y-Y'|],$$

*where $X'$ and $Y'$ are independent and identical copies of $X$ and $Y$, respectively.*

[23] also show that energy distance is equivalent to the weighted $L_2$-distance of the characteristic functions.

**Proposition 2.2.** *[23, Proposition 1] Suppose $X, Y \in \mathbb{R}^p$ are two real-valued independent random variables with finite means and $X'$ and $Y'$ are independent identical copies of $X$ and $Y$. Let $\tilde{f}_X(\cdot)$ and $\tilde{f}_Y(\cdot)$ denote the characteristic function of $X$ and $Y$, respectively, we have*

$$\frac{1}{c_p}\int_{\mathbb{R}^p} \frac{|\tilde{f}_X(t) - \tilde{f}_Y(t)|^2}{|t|^2} dt = 2\mathbb{E}[|X-Y|] - \mathbb{E}[|X-X'|] - \mathbb{E}[|Y-Y'|] = \mathcal{E}(X,Y).$$

*Thus, $\mathcal{E}(X,Y) \geq 0$ with equality to zero if and only if $X$ and $Y$ are identically distributed, i.e., $\tilde{f}_X \equiv \tilde{f}_Y$.*

Suppose that we observe samples $X_1, \ldots, X_n \overset{i.i.d.}{\sim} F$ and $Y_1, \ldots, Y_m \overset{i.i.d.}{\sim} G$, the energy statistics is usually defined as follows (see [23], (6.1)).

$$\frac{2}{nm}\sum_{i=1}^{n}\sum_{j=1}^{m}|X_i - Y_j| - \frac{1}{n^2}\sum_{i=1}^{n}\sum_{j=1}^{n}|X_i - X_j| - \frac{1}{m^2}\sum_{i=1}^{m}\sum_{j=1}^{m}|Y_i - Y_j|.$$

However, above estimator is NOT an unbiased estimator of $\mathcal{E}(X,Y)$. To mitigate this issue, let $h(X_1, X_2, Y_1, Y_2) = \frac{1}{2}\sum_{i=1}^{2}\sum_{j=1}^{2}|X_i - Y_j| - |X_1 - X_2| - |Y_1 - Y_2|$ be a two-sample kernel (see [24, Chapter 12.2]), which an unbiased estimator, i.e., $\mathbb{E}[h] = \mathcal{E}(X,Y)$, then it is easy to verify that

$$\begin{aligned}
&\frac{1}{\binom{n}{2}\binom{m}{2}}\sum_{i_1<i_2, j_1<j_2} h(X_{i_1}, X_{i_2}, Y_{j_1}, Y_{j_2}) \\
=&\frac{2}{nm}\sum_{i=1}^{n}\sum_{j=1}^{m}|X_i - Y_j| - \frac{1}{n(n-1)}\sum_{i,j=1, i\neq j}^{n}|X_i - X_j| - \frac{1}{m(m-1)}\sum_{i,j=1, i\neq j}^{m}|Y_i - Y_j|.
\end{aligned}$$

is a U-statistic and an unbiased estimator of $\mathcal{E}(X,Y)$. Thus, we will use the following definition of energy statistics throughout this paper.

**Definition 2.3** (Unbiased Energy Statistics)**.** *Given samples $X_1, \ldots, X_n \overset{i.i.d.}{\sim} F$ and $Y_1, \ldots, Y_m \overset{i.i.d.}{\sim} G$, the energy statistics between $X$ and $Y$ could be defined as*

$$\mathcal{E}_{n,m}(X,Y) = \frac{2}{nm}\sum_{i=1}^{n}\sum_{j=1}^{m}|X_i - Y_j| - \frac{1}{n(n-1)}\sum_{i,j=1, i\neq j}^{n}|X_i - X_j| - \frac{1}{m(m-1)}\sum_{i,j=1, i\neq j}^{m}|Y_i - Y_j|.$$

# 3 Efficient Computational Method for Energy Statistics

In this section, we will describe the efficient algorithms for energy statistics of both univariate and multivariate random variables in Section 3.1 and Section 3.2, respectively. We will also propose two different methods based on the efficient algorithm of multivariate random variables for two-sample test in Section 3.3.

## 3.1 A Fast Algorithm for Univariate Random Variables

We will start with the fast algorithm for univariate random variables. Let us recall the definition of energy statistics first. Given univariate random variables $X_1, \ldots, X_n \in \mathbb{R}$ and $Y_1, \ldots, Y_m \in \mathbb{R}$, the energy statistic of $X$ and $Y$ is defined below:

$$\mathcal{E}_{n,m}(X,Y) = \frac{2}{nm}\sum_{i=1}^{n}\sum_{j=1}^{m}|X_i - Y_j| - \frac{1}{n(n-1)}\sum_{i,j=1, i\neq j}^{n}|X_i - X_j| - \frac{1}{m(m-1)}\sum_{i,j=1, i\neq j}^{m}|Y_i - Y_j|.$$

For simplicity of notation, we denote above term with $\mathcal{E}_{n,m}$. The following algorithm can compute $\mathcal{E}_{n,m}$ with an average order of complexity $O(N \log N)$, where $N = n + m$. The main idea of this algorithm is sorting the observations first and use a linear-time algorithm to compute the energy statistic with sorted observations.

(1) Sort $X_i$'s and $Y_j$'s, so that we have order statistics $X_{(1)} \leq X_{(2)} \leq \cdots \leq X_{(n)}$ and $Y_{(1)} \leq Y_{(2)} \leq \cdots \leq Y_{(m)}$.

(2) Compute the second term of $\mathcal{E}_{n,m}$ as follows:

$$E_2 = \frac{2}{n(n-1)} \sum_{i=1}^{n-1} i(n-i) \left| X_{(i+1)} - X_{(i)} \right|.$$

(3) Compute the third term of $\mathcal{E}_{n,m}$ as follows:

$$E_3 = \frac{2}{m(m-1)} \sum_{i=1}^{m-1} i(m-i) \left| Y_{(i+1)} - Y_{(i)} \right|.$$

(4) In this step, we will compute the first term of $\mathcal{E}_{n,m}$.

   (a) Merge two ordered series $X_{(1)} \leq X_{(2)} \leq \cdots \leq X_{(n)}$ and $Y_{(1)} \leq Y_{(2)} \leq \cdots \leq Y_{(m)}$ into a single ordered series $Z_{(1)} \leq Z_{(2)} \leq \cdots \leq Z_{(n+m)}$, where each $Z_{(k)}$ is either from $X_{(i)}$'s or from $Y_{(j)}$'s. At the same time, one can generate a sequence $I_i, i = 1, 2, \ldots, n+m$, where $I_i$ records the size of the subset of $Z_{(1)}$ through $Z_{(i)}$ that are from $X_{(i)}$'s.

   (b) Compute the first term of $\mathcal{E}_{n,m}$,

$$E_1 = \frac{2}{nm} \sum_{i=1}^{n+m-1} \left[ I_i(m - i + I_i) + (i - I_i)(n - I_i) \right] \left| Z_{(i+1)} - Z_{(i)} \right|.$$

(5) Compute the energy statistic,

$$\mathcal{E}_{n,m} = E_1 - E_2 - E_3.$$

A stand-alone description of above algorithm can be found in Algorithm 1 of Appendix A. Our result could be summarized in the following theorem.

**Theorem 3.1.** *Given univariate random variables $X_1, \ldots, X_n \in \mathbb{R}$ and $Y_1, \ldots, Y_m \in \mathbb{R}$, there exists an algorithm with complexity $O(N \log N)$, where $N = n + m$, for computing the energy statistic defined in Definition 2.3.*

See Appendix B.1 for the proof and detailed explanations.

### 3.2 A Fast Algorithm for Multivariate Random Variables

In this part, we will introduce a fast algorithm for the energy statistics of multivariate random variables. We will show later in Theorem 4.9 that the estimator produced by this algorithm converges fairly fast. The main idea works as follows: first, projecting the multivariate observations along some random directions; then, using the fast algorithm described in Section 3.1 to compute the energy statistics of projections; last, averaging those energy statistics from different projecting directions.

Formally, suppose we have observations $X_1, \ldots, X_n \in \mathbb{R}^p$ and $Y_1, \ldots, Y_m \in \mathbb{R}^p$ and let $K$ denote the pre-determined number of random projections, the algorithm is as follows:

(1) For each $k$ ($1 \leq k \leq K$), randomly generate projecting direction $u_k$ from Uniform($\mathcal{S}_p$), where $\mathcal{S}_p$ is the unit sphere in $\mathbb{R}^p$.

(2) Let $u_k^T X$ and $u_k^T Y$ denote the projections of $X$ and $Y$. That is,

$$u_k^T X = (u_k^T X_1, \ldots, u_k^T X_n), \text{ and } u_k^T Y = (u_k^T Y_1, \ldots, u_k^T Y_m).$$

   Note that $u_k^T X$ and $u_k^T Y$ are now univariate.

(3) Utilize the fast algorithm described in Section 3.1 to compute the energy statistic of $u_k^T X$ and $u_k^T Y$. Formally, we denote

$$\mathcal{E}_{n,m}^{(k)} = C_p \mathcal{E}_{n,m}(u_k^T X, u_k^T Y),$$

where $C_p$ is the constant defined at the end of Section 1.

(4) Repeat above steps for $K$ times. The final estimator is

$$\overline{\mathcal{E}}_{n,m} = \frac{1}{K} \sum_{k=1}^{K} \mathcal{E}_{n,m}^{(k)},$$

which is refered as Randomly Projected Energy Statistics (RPES). To emphasize the dependency of the above quantity with number of random projections $K$, we sometimes use another notation $\overline{\mathcal{E}}_{n,m;K} \triangleq \overline{\mathcal{E}}_{n,m}$.

A stand-alone description of above algorithm can be found in Algorithm 2 of Appendix A. The following theorem summarizes above result.

**Theorem 3.2.** *For multivariate random variables $X_1, \dots, X_n \in \mathbb{R}$ and $Y_1, \dots, Y_m \in \mathbb{R}$, there exists an algorithm with complexity $O(KN \log N)$, where $N = n + m$, for computing aforementioned $\overline{\mathcal{E}}_{n,m}$, where $K$ is a pre-determined number of random projections.*

We omit the proof since above theorem is a straight-forward conclusion from Theorem 3.1.

### 3.3 Two-Sample Test based on Randomly Projected Energy Statistics (RPES)

The randomly projected energy statistic $\overline{\mathcal{E}}_{n,m}$ could be applied in the two-sample test. Let us recall that we would like to test the null hypotheses $\mathcal{H}_0$ — $X$ and $Y$ are identically distributed — against its alternative. The threshold of the test statistic could be determined by either permutation or the Gamma approximation of asymptotic distribution. Let us recall that we observe $X_1, \dots, X_n \in \mathbb{R}^p$ and $Y_1, \dots, Y_m \in \mathbb{R}^p$. Let $Z = (Z_1, \dots, Z_{n+m}) = (X_1, \dots, X_n, Y_1, \dots, Y_m)$ denote the collection of all observations. Let $\overline{\mathcal{E}}_{n,m}$ denote the proposed estimator defined in Section 3.2. Suppose $\alpha_s$ is the pre-specified significance level of the test and $L$ is the pre-determined number of permutations. The following algorithm describes a two-sample test using permutation to generate the threshold.

(1) For each $l$, $1 \le l \le L$, generate a random permutation of observations: let

$$(X^{\star,l}, Y^{\star,l}) = (X_1^{\star,l}, \dots, X_n^{\star,l}, Y_1^{\star,l}, \dots, Y_m^{\star,l})$$

be a random permutation of $(Z_1, \dots, Z_{n+m})$.

(2) Using the algorithm in Section 3.2, we compute the estimator for $X^{\star,l}$ and $Y^{\star,l}$: $D^{(l)} = \overline{\mathcal{E}}_{n,m}(X^{\star,l}, Y^{\star,l})$. Note that under null hypotheses, $X^{\star,l}$ and $Y^{\star,l}$ are identically distributed.

(3) Reject null hypotheses $\mathcal{H}_0$ if and only if

$$\frac{1 + \sum_{l=1}^{L} I(\overline{\mathcal{E}}_{n,m} > D^{(l)})}{1 + L} > \alpha_s.$$

See Algorithm 3 of Appendix A for a stand-alone description of above algorithm.

We can also find the threshold for test statistic based on the Gamma approximation of its asymptotic distribution. Let $K$ denote the pre-determined number of random projections. The algorithm is as follows:

(1) For each $k$, $1 \leq k \leq K$, randomly generate $u_k$ independently from Unif($\mathcal{S}^{p-1}$).

(2) Use the univariate fast algorithm in Section 3.1 to compute the following quantities:

$$\mathcal{E}_{n,m}^{(k)} = C_p \mathcal{E}_{n,m}(u_k^T X, u_k^T Y),$$
$$S_{1;n,m}^{(k)} = C_p \binom{n+m}{2}^{-1} \sum_{i<j}^{n} |u^T(Z_i - Z_j)|,$$

where constant $C_p$ has been defined at the end of Section 1.

(3) Use the univariate fast algorithm for distance covariance in [10] to compute:

$$S_{2;n,m}^{(k)} = C_p^2 \text{SDC}(u_k^T Z, u_k^T Z),$$

where SDC stands for Sample Distance Covariance defined in [10, eq (3.3)]. Randomly generate $v_k$ from Unif($\mathcal{S}^{p-1}$) and use aforementioned algorithm to compute

$$S_{3;n,m}^{(k)} = C_p^2 \text{SDC}(u_k^T Z, v_k^T Z).$$

(4) Repeat above steps for $k = 1, \ldots, K$ and aggregate the results as follows:

$$\overline{\mathcal{E}}_{n,m} = \frac{1}{K} \sum_{k=1}^{K} \mathcal{E}_{n,m}^{(k)}, \qquad \overline{S}_{1;n,m} = \frac{1}{K} \sum_{k=1}^{K} S_{1;n,m}^{(k)},$$
$$\overline{S}_{2;n,m} = \frac{1}{K} \sum_{k=1}^{K} S_{2;n,m}^{(k)}, \qquad \overline{S}_{3;n,m} = \frac{1}{K} \sum_{k=1}^{K} S_{3;n,m}^{(k)},$$
$$\hat{\alpha} = \frac{1}{2} \frac{\overline{S}_{1;n,m}^2}{\frac{1}{K}\overline{S}_{2;n,m} + \frac{K-1}{K}\overline{S}_{3;n,m}}, \tag{3.1}$$
$$\hat{\beta} = \frac{1}{2} \frac{\overline{S}_{1;n,m}}{\frac{1}{K}\overline{S}_{2;n,m} + \frac{K-1}{K}\overline{S}_{3;n,m}}. \tag{3.2}$$

(5) Reject null hypothses $\mathcal{H}_0$ if and only if $(n+m)\overline{\mathcal{E}}_{n,m} + \overline{S}_{1;n,m} > \text{Gamma}(1-\alpha_s; \hat{\alpha}, \hat{\beta})$, where $\text{Gamma}(1-\alpha_s; \hat{\alpha}, \hat{\beta})$ is the $1-\alpha_s$ percentile of Gamma distribution with shape parameter $\hat{\alpha}$ and rate parameter $\hat{\beta}$; Otherwise, accept it.

See Algorithm 4 of Appendix A for a stand-alone description of above algorithm.

# 4 Theoretical Properties

Firstly, we will show some nice properties of random projections in energy distance and energy statistics in Section 4.1. Then, we will study the asymptotic properties of energy statistics $\mathcal{E}_{n,m}$ and randomly projected energy statistics $\overline{\mathcal{E}}_{n,m}$ in Section 4.2 and 4.3, respectively.

## 4.1 Properties of Random Projections in Energy Distance

We will study some properties of randomly projected energy distance and energy statistics in this part. We begin a sufficient and necessary condition of equality of distributions.

**Lemma 4.1.** *Suppose $u$ is some random point on unit sphere $\mathcal{S}^{p-1}$: $u \in \mathcal{S}^{p-1} := \{u \in \mathbb{R}^p : |u| = 1\}$. We have*

*random vector $X \in \mathbb{R}^p$ has the same distribution with random vector $Y \in \mathbb{R}^p$*

*if and only if*

$$\mathcal{E}(u^T X, u^T Y) = 0 \text{ for any } u \in \mathcal{S}^{p-1}.$$

The following result allows us to regard energy distance / energy statistics of multivariate random variables as the integration of energy distance / energy statistics of univariate random variables. This result provides the foundation of our proposed method in Section 3.2.

**Lemma 4.2.** *Suppose $u$ is some random point on unit sphere $\mathcal{S}^{p-1}$. Let $\mu$ denote the uniform probability measure on $\mathcal{S}^{p-1}$. Then, for random vectors $X, Y \in \mathbb{R}^p$ with $\mathbb{E}[|X|] < \infty, \mathbb{E}[|Y|] < \infty$, we have*

$$\mathcal{E}(X, Y) = C_p \int_{\mathcal{S}^{p-1}} \mathcal{E}(u^T X, u^T Y) d\mu(u),$$

*where $C_p$ is the constant defined at the end of Section 1. Similarly, for energy statistics, we have*

$$\mathcal{E}_{n,m}(X, Y) = C_p \int_{\mathcal{S}^{p-1}} \mathcal{E}_{n,m}(u^T X, u^T Y) d\mu(u).$$

## 4.2 Asymptotic Properties of Energy Statistics $\mathcal{E}_{n,m}$

As showed in Section 2, the energy statistics $\mathcal{E}_{n,m}$ is a two-sample u-statistics with respect to kernel

$$h(X_1, X_2, Y_1, Y_2) = \frac{1}{2} \sum_{i=1}^{2} \sum_{j=1}^{2} |X_i - Y_j| - |X_1 - X_2| - |Y_1 - Y_2|$$

which is a two-sample kernel. Before analyzing the asymptotic properties of $\mathcal{E}_{n,m}$, let us define the following quantities that will play important roles in subsequent studies:

$$\begin{aligned}
h_{10} &= h_{10}(X_1) = \mathbb{E}_{X_2,Y_1,Y_2}[h(X_1, X_2, Y_1, Y_2)], \\
h_{01} &= h_{01}(Y_1) = \mathbb{E}_{X_1,X_2,Y_2}[h(X_1, X_2, Y_1, Y_2)], \\
h_{20} &= h_{20}(X_1, X_2) = \mathbb{E}_{Y_1,Y_2}[h(X_1, X_2, Y_1, Y_2)], \\
h_{02} &= h_{02}(Y_1, Y_2) = \mathbb{E}_{X_1,X_2}[h(X_1, X_2, Y_1, Y_2)], \\
h_{11} &= h_{11}(X_1, Y_1) = \mathbb{E}_{X_2,Y_2}[h(X_1, X_2, Y_1, Y_2)],
\end{aligned}$$

where the two subindexes represent how many $X$'s and $Y$'s in the functions, respectively.

**Lemma 4.3** (Generic Formula)**.** *If $\mathbb{E}[|X|] + \mathbb{E}[|Y|] < \infty$, for independent $X_1$, $X_2$, $X$, $X'$, $Y_1$, $Y_2$, $Y$ and*

$Y'$, w we have

$$h_{10}(X_1) = \mathbb{E}_Y[|X_1 - Y|] + \mathbb{E}_{X,Y}[|X - Y|] - \mathbb{E}_X[|X_1 - X|] - \mathbb{E}_{Y,Y'}[|Y - Y'|], \tag{4.3}$$
$$h_{01}(Y_1) = \mathbb{E}_X[|X - Y_1|] + \mathbb{E}_{X,Y}[|X - Y|] - \mathbb{E}_{X,X'}[|X - X'|] - \mathbb{E}_Y[|Y_1 - Y|], \tag{4.4}$$
$$h_{20}(X_1, X_2) = \mathbb{E}_Y[|X_1 - Y|] + \mathbb{E}_Y[|X_2 - Y|] - |X_1 - X_2| - \mathbb{E}_{Y,Y'}[|Y - Y'|], \tag{4.5}$$
$$h_{02}(Y_1, Y_2) = \mathbb{E}_X[|X - Y_1|] + \mathbb{E}_X[|X - Y_1|] - |Y_1 - Y_2| - \mathbb{E}_{X,X'}[|X - X'|], \tag{4.6}$$
$$h_{11}(X_1, Y_1) = \frac{1}{2}|X_1 - Y_1| + \frac{1}{2}\mathbb{E}_X[|X - Y_1|] + \frac{1}{2}\mathbb{E}_Y[|X_1 - Y|] + \frac{1}{2}\mathbb{E}_{X,Y}[|X - Y|] - \mathbb{E}_X[|X_1 - X|] - \mathbb{E}_Y[|Y_1 - Y|]. \tag{4.7}$$

We can also define $h_{21}$, $h_{12}$ and $h_{22}$ in a similar way but we do not list them here as they are not important in subsequent analysis. The corresponding variance of $h_{i,j}$ is denoted by

$$\sigma_{ij}^2 = \text{Var}[h_{ij}], 1 \le i + j \le 2, 1 \le i, j \le 2.$$

Then, by the result [14] Section 2.2 Theorem 2, the variance of $\mathcal{E}_{n,m}$ can be represented as follows.

**Lemma 4.4** (Variance of two-sample U-statistics). *Suppose Var*$[h(X_1, X_2, Y_1, Y_2)] < \infty$ *and* $n, m \ge 4$, *then the variance* $\mathcal{E}_{n,m}(X, Y)$ *is*

$$\begin{aligned} Var[\mathcal{E}_{n,m}] &= \frac{1}{\binom{n}{2}\binom{m}{2}} \sum_{i,j=0, i+j\ge 1}^{2} \binom{2}{i}\binom{2}{j}\binom{n-2}{2-i}\binom{m-2}{2-j}\sigma_{ij}^2 \\ &= \frac{4}{m}\sigma_{01}^2 + \frac{4}{n}\sigma_{10}^2 + (\frac{16}{mn} + \frac{1}{m^2})\sigma_{01}^2 + (\frac{16}{mn} + \frac{1}{n^2})\sigma_{10}^2 \\ &\quad + \frac{2}{m^2}\sigma_{02}^2 + \frac{2}{n^2}\sigma_{20}^2 + \frac{16}{nm}\sigma_{11}^2 + O(\frac{1}{n^2 m}) + O(\frac{1}{nm^2}) \end{aligned}$$

[14] also shows that $\mathcal{E}_{n,m}$ is asymptotically normal under mild conditions.

**Theorem 4.5.** *([14, Section 3.7, Theorem 1]) Let* $N = n + m$ *denote the total number of observations. Suppose there exists constant* $0 < \eta < 1$ *such that* $n/N \to \eta$ *and* $m/N \to 1 - \eta$ *as* $n, m \to \infty$. *If Var*$[h(X_1, X_2, Y_1, Y_2)] < \infty$ *and* $\sigma_{10}^2 + \sigma_{01}^2 > 0$, *then* $\sqrt{N}(\mathcal{E}_{n,m} - \mathcal{E})$ *converges in distribution to a normal distribution with mean zero and variance* $4\sigma_{10}^2/\eta + 4\sigma_{01}^2/(1-\eta)$, *i.e.,*

$$\sqrt{N}(\mathcal{E}_{n,m} - \mathcal{E}) \xrightarrow{D} \mathcal{N}(0, 4\sigma_{10}^2/\eta + 4\sigma_{01}^2/(1-\eta)),$$

*where* $\mathcal{E}$ *is the energy distance* $\mathcal{E} = \mathbb{E}[\mathcal{E}_{n,m}]$.

Now, we assume that $X$ has the same distribution with $Y$. Then, the formulas of $h_{ij}$ could be simplified.

**Lemma 4.6.** *If* $X$ *and* $Y$ *are identically distributed, then we have*

$$h_{10}(X_1) = 0, \qquad h_{01}(Y_1) = 0, \tag{4.8}$$
$$h_{20}(X_1, X_2) = \mathbb{E}_X[|X_1 - X|] + \mathbb{E}_X[|X_2 - X|] - |X_1 - X_2| - \mathbb{E}_{X,X'}[|X - X'|], \tag{4.9}$$
$$h_{02}(Y_1, Y_2) = \mathbb{E}_Y[|Y_1 - Y|] + \mathbb{E}_Y[|Y_2 - Y|] - |Y_1 - Y_2| - \mathbb{E}_{Y,Y'}[|Y - Y'|], \tag{4.10}$$
$$h_{11}(X_1, Y_1) = \frac{1}{2}\left(|X_1 - Y_1| - \mathbb{E}_X[|X_1 - X|] - \mathbb{E}_X[|Y_1 - X|] + \mathbb{E}_{X,X'}[|X - X'|]\right). \tag{4.11}$$

The proof of this lemma is straightforward by noting the fact that the usage of $X$ and $Y$ is interchangeable as they are identically independently distributed.

When $X$ has the same distribution with $Y$, $\mathcal{E}_{n,m}$ is no longer asymptotically normal. Instead, $(n+m)\mathcal{E}_{n,m}$ converges to a sum of (possibly infinite) independent chi-squared random variables.

**Theorem 4.7.** *Let $N = n + m$ denote the total number of observations. Suppose there exists constant $0 < \eta < 1$ such that $n/N \to \eta$ and $m/N \to 1 - \eta$ as $n, m \to \infty$. If $X$ and $Y$ are identically distributed, the asymptotic distribution of $\mathcal{E}_{n,m}$ is*

$$N\mathcal{E}_{n,m} \xrightarrow{D} \sum_{l=1}^{\infty} \frac{\lambda_l}{\eta(1-\eta)}(Z_l^2 - 1),$$

*where $Z_1, Z_2, \ldots$ are independent standard normal random variables and $\lambda_l$'s are defined in Lemma B.2 and*

$$\sum_{l=1}^{\infty} \lambda_l = \mathbb{E}_{X,X'}[|X - X'|], \qquad \sum_{l=1}^{\infty} \lambda_l^2 = DC(X, X),$$

*where $DC(X, X)$ is the distance covariance of $X$, see [9].*

See appendix for a proof.

## 4.3 Asymptotic Properties of Randomly Projected Energy Statistics $\overline{\mathcal{E}}_{n,m}$

Let us recall some notations. The randomly projected energy statistics $\overline{\mathcal{E}}_{n,m}$ is defined as

$$\overline{\mathcal{E}}_{n,m} = \frac{1}{K}\sum_{k=1}^{K} \mathcal{E}_{n,m}^{(k)} = \frac{1}{K}\sum_{k=1}^{K} C_p \mathcal{E}_{n,m}(u_k^T X, u_k^T Y),$$

where constant $C_p$ has been defined at the end of Section 1 and $u_k$'s are independent samples from $\text{Unif}(\mathcal{S}^{p-1})$. Note that $\mathcal{E}_{n,m}(u_k^T X, u_k^T Y)$ is a U-statistic for any $k$ and $\overline{\mathcal{E}}_{n,m}$ is also a U-statistic as

$$\begin{aligned}
\overline{\mathcal{E}}_{n,m} &= \frac{1}{K}\sum_{k=1}^{K} \frac{C_p}{\binom{n}{2}\binom{m}{2}} \sum_{i_1<i_2, j_1<j_2} h(u_k^T X_{i_1}, u_k^T X_{i_2}, u_k^T Y_{j_1}, u_k^T Y_{j_2}) \\
&= \frac{1}{\binom{n}{2}\binom{m}{2}} \sum_{i_1<i_2, j_1<j_2} \frac{1}{K}\sum_{k=1}^{K} C_p h(u_k^T X_{i_1}, u_k^T X_{i_2}, u_k^T Y_{j_1}, u_k^T Y_{j_2}) \\
&\triangleq \frac{1}{\binom{n}{2}\binom{m}{2}} \sum_{i_1<i_2, j_1<j_2} \overline{h}(X_{i_1}, X_{i_2}, Y_{j_1}, Y_{j_2}),
\end{aligned}$$

where

$$\overline{h}(X_{i_1}, X_{i_2}, Y_{j_1}, Y_{j_2}) = \frac{1}{K}\sum_{k=1}^{K} C_p h(u_k^T X_{i_1}, u_k^T X_{i_2}, u_k^T Y_{j_1}, u_k^T Y_{j_2})$$

is the kernel of $\overline{\mathcal{E}}_{n,m}$. Let us define the following notations that will be essential in analyzing the asymptotic properties of $\overline{\mathcal{E}}_{n,m}$:

$$\begin{aligned}
\overline{h}_{10} &= \overline{h}_{10}(X_1) = \mathbb{E}_{X_2,Y_1,Y_2}[\overline{h}(X_1,X_2,Y_1,Y_2)],\\
\overline{h}_{01} &= \overline{h}_{01}(Y_1) = \mathbb{E}_{X_1,X_2,Y_2}[\overline{h}(X_1,X_2,Y_1,Y_2)],\\
\overline{h}_{20} &= \overline{h}_{20}(X_1,X_2) = \mathbb{E}_{Y_1,Y_2}[\overline{h}(X_1,X_2,Y_1,Y_2)],\\
\overline{h}_{02} &= \overline{h}_{02}(Y_1,Y_2) = \mathbb{E}_{X_1,X_2}[\overline{h}(X_1,X_2,Y_1,Y_2)],\\
\overline{h}_{11} &= \overline{h}_{11}(X_1,Y_1) = \mathbb{E}_{X_2,Y_2}[\overline{h}(X_1,X_2,Y_1,Y_2)],
\end{aligned}$$

where the expecations are taken with respect to $(X,Y)$ given random projections $U$. We also let $\overline{\sigma}_{ij}^2$ denote the conditional variance of $\overline{h}_{ij}$ given all projection directions $U=(u_1,\dots,u_K)$,

$$\overline{\sigma}_{ij}^2 = \overline{\sigma}_{ij}^2(U) = \text{Var}_{X,Y}[\overline{h}_{ij}|U].$$

### 4.3.1 Asymptotic Properties in Inequality of Distribution

By Lemma 4.4 and the Law of Total Variance, we have the following result on the variance of $\overline{\mathcal{E}}_{n,m}$.

**Lemma 4.8** (Variance of $\overline{\mathcal{E}}_{n,m}$). *Suppose $Var[h(X_1,X_2,Y_1,Y_2)] < \infty$ and $n,m \geq 4$, then the variance $\overline{\mathcal{E}}_{n,m}$ is*

$$\begin{aligned}
Var[\overline{\mathcal{E}}_{n,m}] = \frac{1}{K} Var_u\left[\mathcal{E}(u^TX,u^TY)\right] &+ \mathbb{E}_U\left[\frac{4}{m}\overline{\sigma}_{01}^2 + \frac{4}{n}\overline{\sigma}_{10}^2\right]\\
&+ \mathbb{E}_U\left[(\frac{16}{mn}+\frac{1}{m^2})\overline{\sigma}_{01}^2 + (\frac{16}{mn}+\frac{1}{n^2})\overline{\sigma}_{10}^2\right]\\
&+ \mathbb{E}_U\left[\frac{2}{m^2}\overline{\sigma}_{02}^2 + \frac{2}{n^2}\overline{\sigma}_{20}^2 + \frac{16}{nm}\overline{\sigma}_{11}^2\right] + O(\frac{1}{n^2m}) + O(\frac{1}{nm^2}).
\end{aligned}$$

As an immediate result from Lemma 4.8, we have the following theorem on the asymptotic properties of $\overline{\mathcal{E}}_{n,m}$.

**Theorem 4.9.** *Suppose $Var[h(X_1,X_2,Y_1,Y_2)] < \infty$. Let $N = n+m$ and assume $n/N \to \eta$ as $N \to \infty$, where $0 < \eta < 1$, then we have*

$$\overline{\mathcal{E}}_{n,m} \xrightarrow{p} \mathcal{E}(X,Y) \text{ as } N\to\infty, K\to\infty.$$

*The asymptotic distribution of $\overline{\mathcal{E}}_{n,m}$ could differ under different conditions.*

*(1) If $K\to\infty$ and $K/N\to 0$, then*

$$\sqrt{K}(\overline{\mathcal{E}}_{n,m} - \mathcal{E}(X,Y)) \xrightarrow{D} \mathcal{N}(0, Var_u[\mathcal{E}(u^TX,u^TY)]).$$

*(2) If $N\to\infty$ and $K/N\to\infty$, then*

$$\sqrt{N}(\overline{\mathcal{E}}_{n,m} - \mathcal{E}(X,Y)) \xrightarrow{D} \mathcal{N}(0, \frac{4}{\eta}\mathbb{E}_U[\overline{\sigma}_{10}^2] + \frac{4}{1-\eta}\mathbb{E}_U[\overline{\sigma}_{01}^2]).$$

*(3) If $N\to\infty$ and $K/N\to C$, where $0<C<\infty$, then*

$$\sqrt{N}(\overline{\mathcal{E}}_{n,m} - \mathcal{E}(X,Y)) \xrightarrow{D} \mathcal{N}(0, \frac{1}{C}Var_u[\mathcal{E}(u^TX,u^TY)] + \frac{4}{\eta}\mathbb{E}_U[\overline{\sigma}_{10}^2] + \frac{4}{1-\eta}\mathbb{E}_U[\overline{\sigma}_{01}^2]).$$

### 4.3.2 Asymptotic Properties in Equality of Distribution

It is of more interest to study the asymptotic properties of $\overline{\mathcal{E}}_{n,m}$ under the condition that $X$ has the same distribution with $Y$. We have the following lemma under this condition.

**Lemma 4.10.** *If $X$ has the same distribution with $Y$, we have*

$$Var_u[\mathcal{E}(u^T X, u^T Y)] = 0,$$

*and,*

$$\overline{h}_{10} = 0, \overline{h}_{01} = 0 \text{ with probability } 1,$$

*which implies*

$$\overline{\sigma}_{10}^2 = Var[\overline{h}_{10}|U] = 0, \overline{\sigma}_{01}^2 = Var[\overline{h}_{01}|U] = 0.$$

*Therefore, the variance of $\overline{\mathcal{E}}_{n,m}$ could be expressed as*

$$Var[\overline{\mathcal{E}}_{n,m}] = \mathbb{E}_U\left[\frac{2}{m^2}\overline{\sigma}_{02}^2 + \frac{2}{n^2}\overline{\sigma}_{20}^2 + \frac{16}{nm}\overline{\sigma}_{11}^2\right] + O(\frac{1}{n^2 m}) + O(\frac{1}{nm^2}).$$

See appendix for the proof.

We should also be aware of a result, which is similar with Lemma B.2. This result will play an important role for our main theorem and its proof.

**Lemma 4.11.** *The kernel $\overline{\mathbf{k}}(\cdot,\cdot)$ defined as*

$$\overline{\mathbf{k}}(X_1, X_2) = \frac{C_p}{K}\sum_{k=1}^{K}\mathbb{E}_X[|u_k^T(X_1 - X)|] + \mathbb{E}_X[|u_k^T(X_2 - X)|] - |u_k^T(X_1 - X_2)| - \mathbb{E}_{X,X'}[|u_k^T(X - X')|]$$

*is a positive kernel and thus there exists $\overline{\phi}_1(\cdot), \overline{\phi}_2(\cdot), \ldots$ such that*

$$\overline{\mathbf{k}}(X_1, X_2) = \sum_{i=1}^{\infty}\overline{\lambda}_i\overline{\phi}_i(X_1)\overline{\phi}_i(X_2),$$

*where $\overline{\lambda}_1 \geq \overline{\lambda}_2 \geq \ldots \geq 0$, $\mathbb{E}[\phi_i(X)] = 0$, $\mathbb{E}[\overline{\phi}_i(X)^2] = 1$ and $\mathbb{E}[\overline{\phi}_i(X)\overline{\phi}_j(X)] = 0$, $i = 1, 2, \ldots, \infty$, $i \neq j$.*

*Proof.* It is worth noting that $\overline{\mathbf{k}}(\cdot,\cdot)$ a positive kernel as it is the sum of a collection of positive kernel. The rest follows by Mercer's Theorem. □

Equipped with above two lemmas, we can conclude that $\overline{\mathcal{E}}_{n,m}$ also converges to a weighted sum of chi-square random variables when the collection of random projections $U$ is given.

**Theorem 4.12.** *Let $N = n + m$ denote the total number of observations. Suppose there exists constant $0 < \eta < 1$ such that $n/N \to \eta$ and $m/N \to 1 - \eta$ as $n, m \to \infty$. If $X$ and $Y$ are identically distributed and all projection directions $U = (u_1, \ldots, u_K)$ are given, the asymptotic distribution of $\mathcal{E}_{n,m}$ is*

$$N\overline{\mathcal{E}}_{n,m} \xrightarrow{D} \sum_{l=1}^{\infty}\frac{\overline{\lambda}_l}{\eta(1-\eta)}(Z_l^2 - 1) = \frac{1}{\eta(1-\eta)}\sum_{l=1}^{\infty}\overline{\lambda}_l Z_l^2 - \frac{1}{\eta(1-\eta)}\sum_{l=1}^{\infty}\overline{\lambda}_l,$$

*where $Z_1, Z_2, \ldots$ are independent standard normal random variables and $\overline{\lambda}_l$'s are the eigenvalues associated with kernel $\overline{\mathrm{k}}(\cdot,\cdot)$ in Lemma 4.11. We also have*

$$\sum_{l=1}^{\infty} \overline{\lambda}_l = \frac{C_p}{K} \sum_{k=1}^{K} \mathbb{E}_{X,X'}[|u_k^T(X - X')|], \qquad \sum_{l=1}^{\infty} \overline{\lambda}_l^2 = \frac{C_p^2}{K^2} \sum_{k,k'=1}^{K} DC(u_k^T X, u_{k'}^T X),$$

*where $DC(u_k^T X, u_{k'}^T X)$ is the distance covariance between $u_k^T X$ and $u_{k'}^T X$.*

See appendix for the proof.

Usually, $\sum_{l=1}^{\infty} \overline{\lambda}_l Z_l^2$ is a weighted sum of infinite many chi-squared random variables. As a result, there is no close form for the asymptotic distribution of $\overline{\mathcal{E}}_{n,m}$. But, we can approximate it by a gamma distribution with first two moments matched, see [4]. As a result, $\sum_{l=1}^{\infty} \overline{\lambda}_l Z_l^2$ could be approximated by Gamma$(\alpha, \beta)$ with density function

$$\frac{\beta^\alpha}{\Gamma(\alpha)} x^{\alpha-1} e^{-\beta x}, x > 0,$$

where

$$\alpha = \frac{1}{2} \frac{(\sum_{l=1}^{\infty} \overline{\lambda}_l)^2}{\sum_{l=1}^{\infty} \overline{\lambda}_l^2}, \quad \beta = \frac{1}{2} \frac{\sum_{l=1}^{\infty} \overline{\lambda}_l}{\sum_{l=1}^{\infty} \overline{\lambda}_l^2}.$$

The following proposition gives a specific way to approximate $\sum_{l=1}^{\infty} \overline{\lambda}_l$ and $\sum_{l=1}^{\infty} \overline{\lambda}_l^2$ from data.

**Proposition 4.13.** *Let $Z$ denote the collection of all observations,*

$$Z = (Z_1, \ldots, Z_{n+m}) = (X_1, \ldots, X_n, Y_1, \ldots, Y_m).$$

*When $X$ and $Y$ have the same distribution, we can approximate $\sum_{l=1}^{\infty} \overline{\lambda}_l$ and $\sum_{l=1}^{\infty} \overline{\lambda}_l^2$ as follows:*

$$\sum_{l=1}^{\infty} \overline{\lambda}_l \approx \frac{C_p}{K} \sum_{k=1}^{K} \frac{1}{(n+m)(n+m-1)} \sum_{i \neq j}^{n+m} |u_k^T(Z_i - Z_j)|$$

$$\sum_{l=1}^{\infty} \overline{\lambda}_l^2 \approx \frac{C_p^2}{K^2} \sum_{k=1}^{K} SDC(u_k^T Z, u_k^T Z) + \frac{(K-1)C_p^2}{K^2} \sum_{k=1}^{K} SDC(u_k^T Z, v_k^T Z),$$

*where $SDC(\cdot,\cdot)$ denotes the sample distance covariance and $v_1, \ldots, v_K$ are all independent random variables from Unif$(\mathcal{S}^{p-1})$.*

See appendix for the reasoning and justification.

# 5 Simulations

## 5.1 Speed Comparison with Direct Method

In this section, we compare the computing speed of the proposed algorithms for univariate random variables and multivariate random variables with direct method by Definition 2.3. This experiment is run on a laptop (MacBook Pro Retina, 13-inch, Early 2015, 2.7 GHz Intel Core i5, 8 GB 1867 MHz DDR3) with MATLAB R2016b (9.1.0.441655). Figure 1 summarizes the time cost of each method against sample size. Note that the scale of time elapsed is different in each subfigure. The result demonstrates the computational advantage of the fast algorithm when sample size is large.

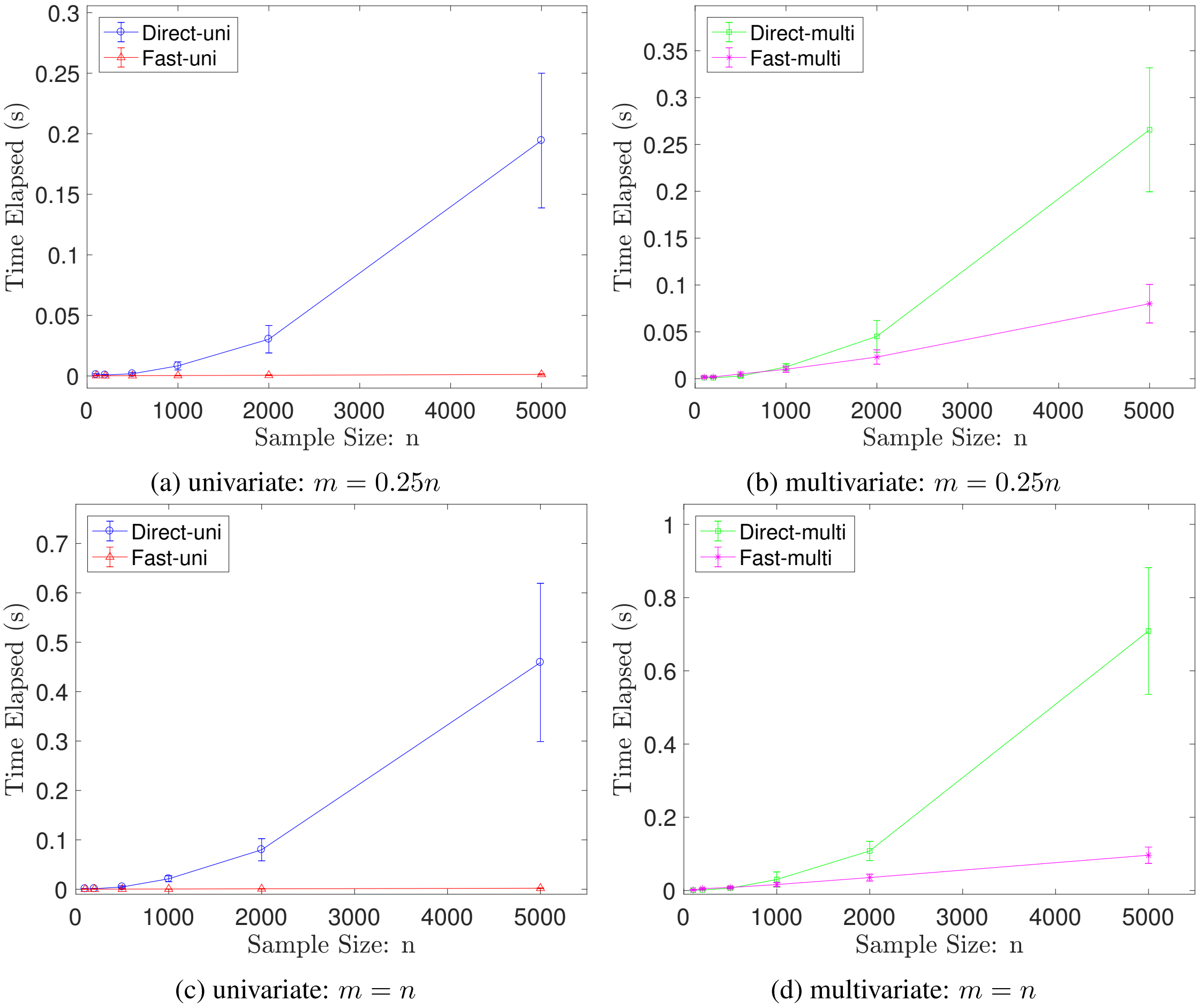

(a) univariate: $m = 0.25n$

(b) multivariate: $m = 0.25n$

(c) univariate: $m = n$

(d) multivariate: $m = n$

Figure 1: Speed Comparison:"Direct-uni" and "Direct-multi" represent the direct method for univariate and multivariate random variables, respectively; "Fast-uni" represents the fast algorithm for univariate random variables described in Section 3.1; "Fast-multi" represents the fast algorithm for multivariate random variables described in Section 3.2 and the number of Monte Carlo iterations is chosen to be $K = 50$. The dimension of the multivariate random varialbes is fixed to be $p = 10$. We let the ratio of sample size of $Y$ over sample size of $X$ be either $0.25$ or $1$. The experiment is repeated for $400$ times.

## 5.2 Impact of Sample Size, Data Dimension and Number of Random Projections

In this section, we will use synthetic data to study the impact of sample size $(n, m)$, data dimension $p$ and Number of Random Projections $K$ on the convergence and test power of multivariate energy statistics. The significance level is set to be $\alpha_s = 0.05$. Each experiment will be repeated for $400$ times to achieve reliable means and variances.

In the following two examples, we will fix sample size $n = 5000, m = 5000$ and let data dimension $p$ vary in $(5, 10, 50, 100, 500)$ and number of random projections $K$ vary in $(10, 50, 100, 500, 1000)$. In Example 5.1, $X$ and $Y$ are identically distributed while they are not in Example 5.2. The result in these two examples suggests that $K = 50$ should suffice when sample size is sufficiently large, regardless of the data dimension.

**Example 5.1.** *We generate random vector $X, Y \sim \mathcal{N}(0, I_p)$, which implies $X$ and $Y$ are identically dis-*

*tributed.*

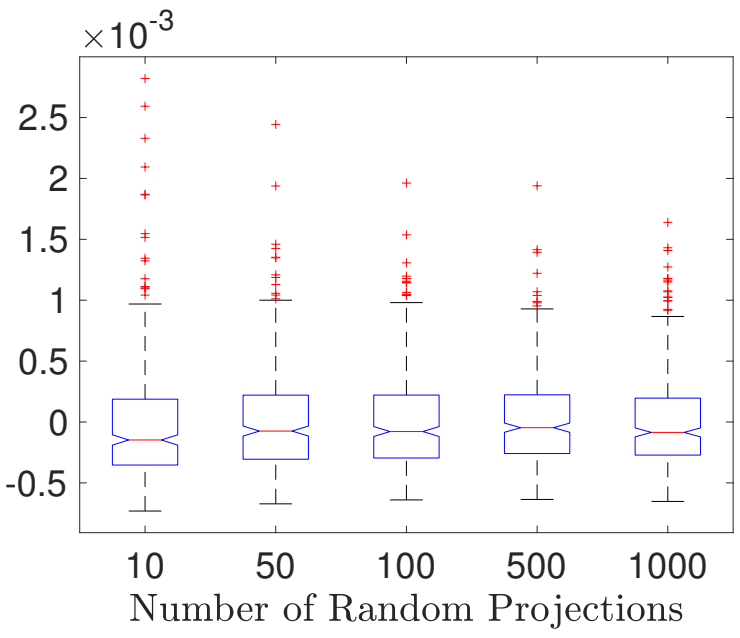

(a) p=5

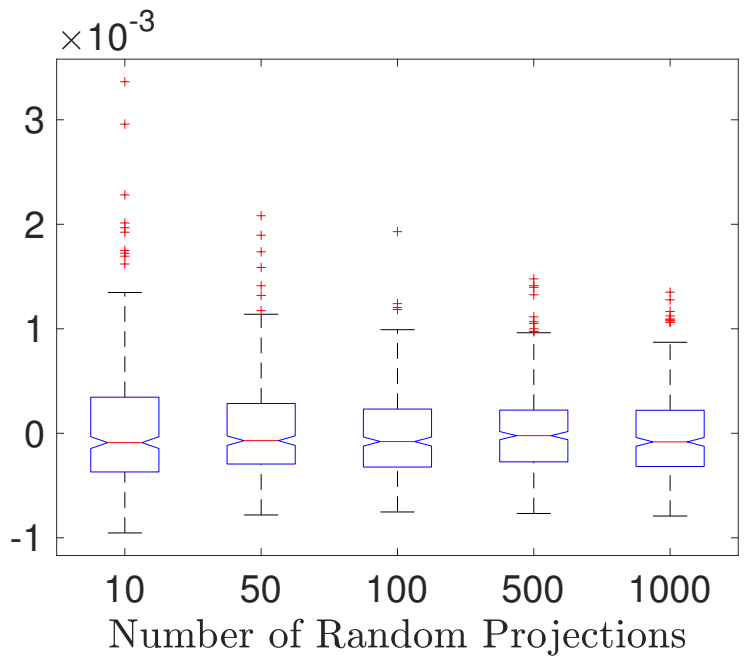

(b) p=10

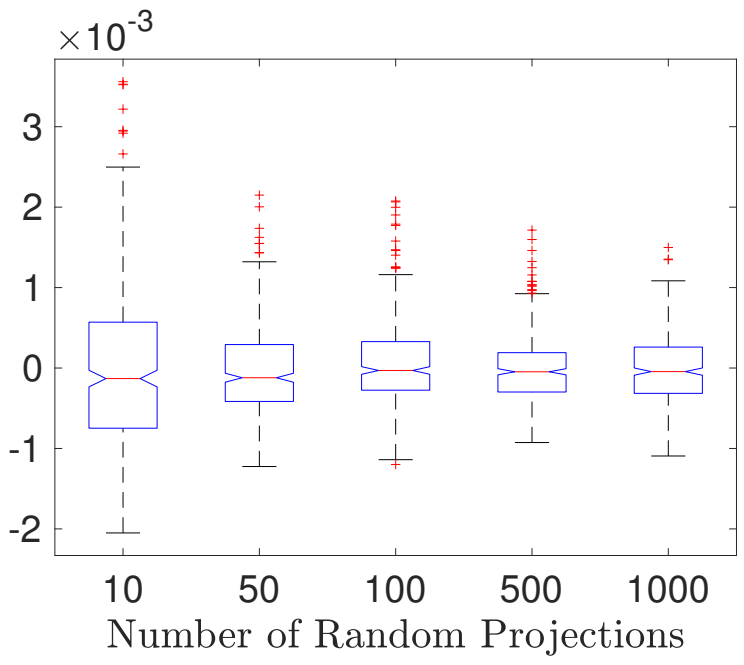

(c) p=50

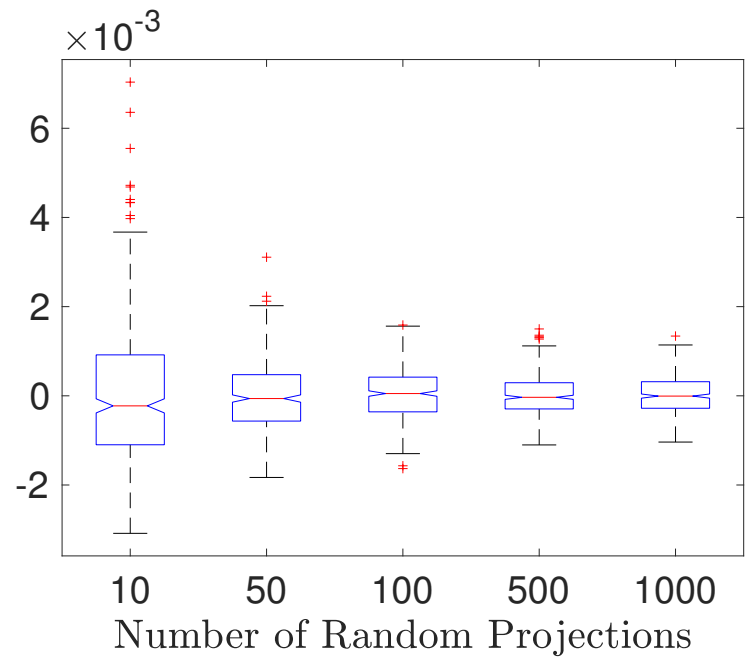

(d) p=100

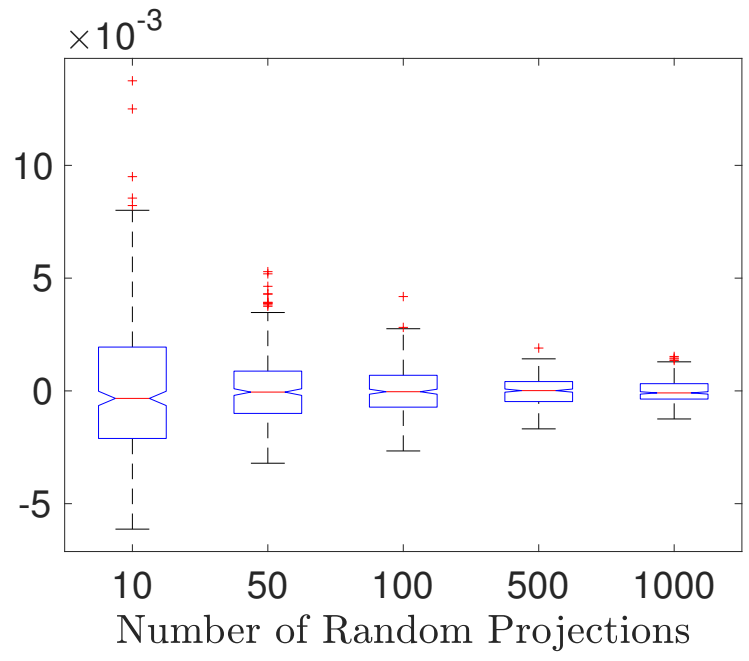

(e) p=500

Figure 2: Boxplots of estimators in Example 5.1. Sample size of $X$ and $Y$ are fixed to be $n = 2000$, $m = 2000$, respectively; the result is based on 400 repeated experiments.

**Example 5.2.** *We generate random vector* $X \sim \mathcal{N}(0, I_p)$*,* $Y \sim t(5)^{(p)}$*, where each entry of* $Y$ *follows t-distribution with degrees of freedom* 5*. In this case, the distribution of* $X$ *is different from the distribution of* $Y$*.*

### 5.3 Compare with Other Two-Sample Tests

We compare our method — Randomly Projected Energy Statistics (RPES) with direct method of Energy Statistics (ES) as well as the most popular alternative in recent literature — the Maximum Mean Discrepancy (MMD) proposed by [8]. Specifically, we use the MMD with Gaussian kernels in our implementation. To obtain reliable estimate of test power, the experiments will be repeated for 200 times.

In the following example, we will measure the power of those tests in distinguishing minor difference in mean of two multivariate normal distribution.

**Example 5.3.** *We generate random vector* $X \sim \mathcal{N}(0, I_p)$*,* $Y \sim \mathcal{N}(\mu, I_p)$*. We let* $\mu = (0.1, 0, \ldots, 0)^t$*, where the first entry of* $\mu$ *is 0.1 while the rest entries are all* 0*.We let* $p = 5$ *and* $p = 50$ *to represent low dimensional case and moderate dimensional case, respectively. In the* $p = 5$ *case, the sample sizes* $n = m$ *is from 500 to 2500 with an increment 100; and in the* $p = 50$ *case, the sample size n is from 500 to 5000 with an increment 250.*

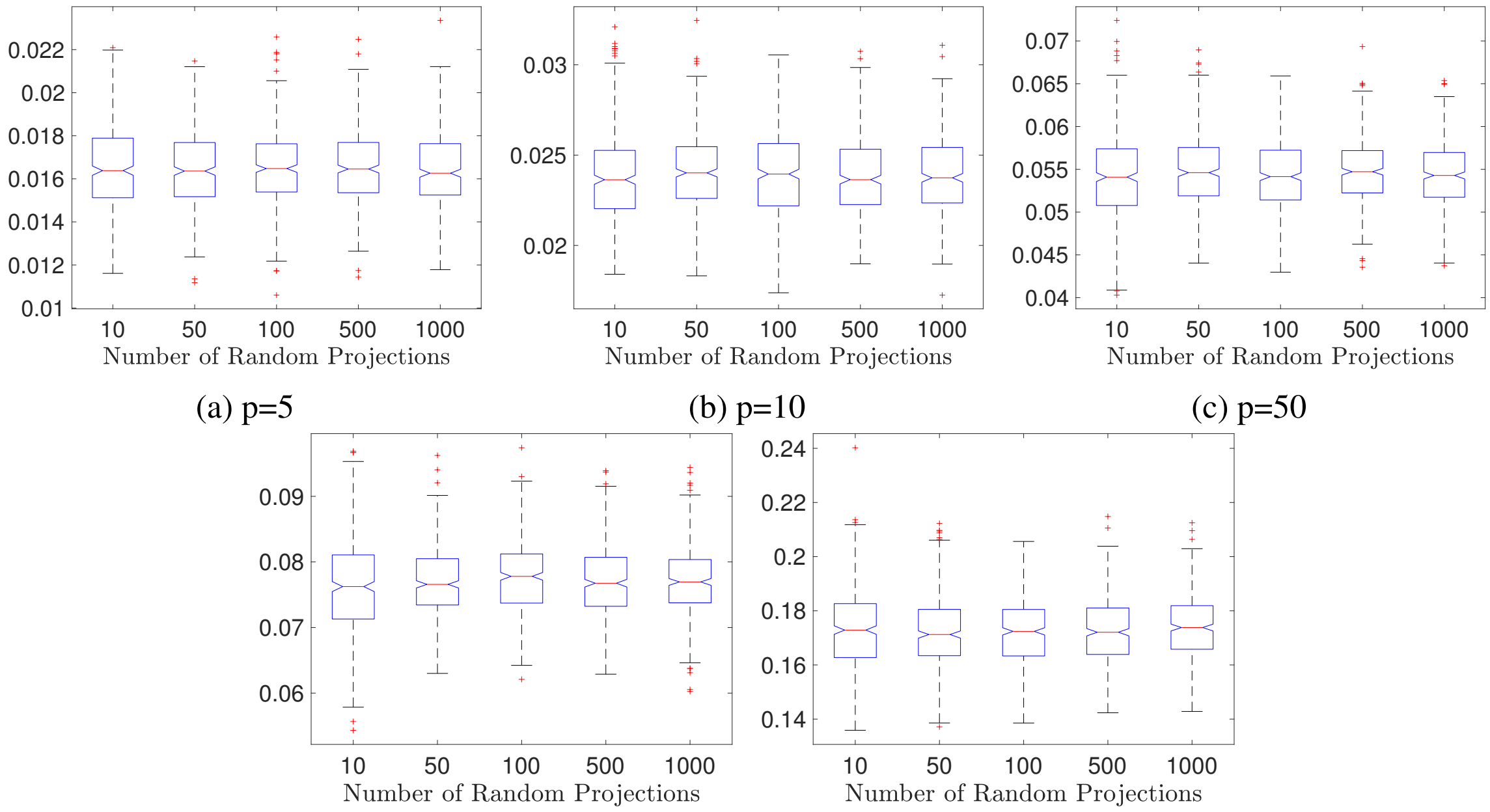

(a) p=5 (b) p=10 (c) p=50

(d) p=100 (e) p=500

Figure 3: Boxplots of estimators in Example 5.2. Sample size of $X$ and $Y$ are fixed to be $n = 2000$, $m = 2000$, respectively; the result is based on $400$ repeated experiments.

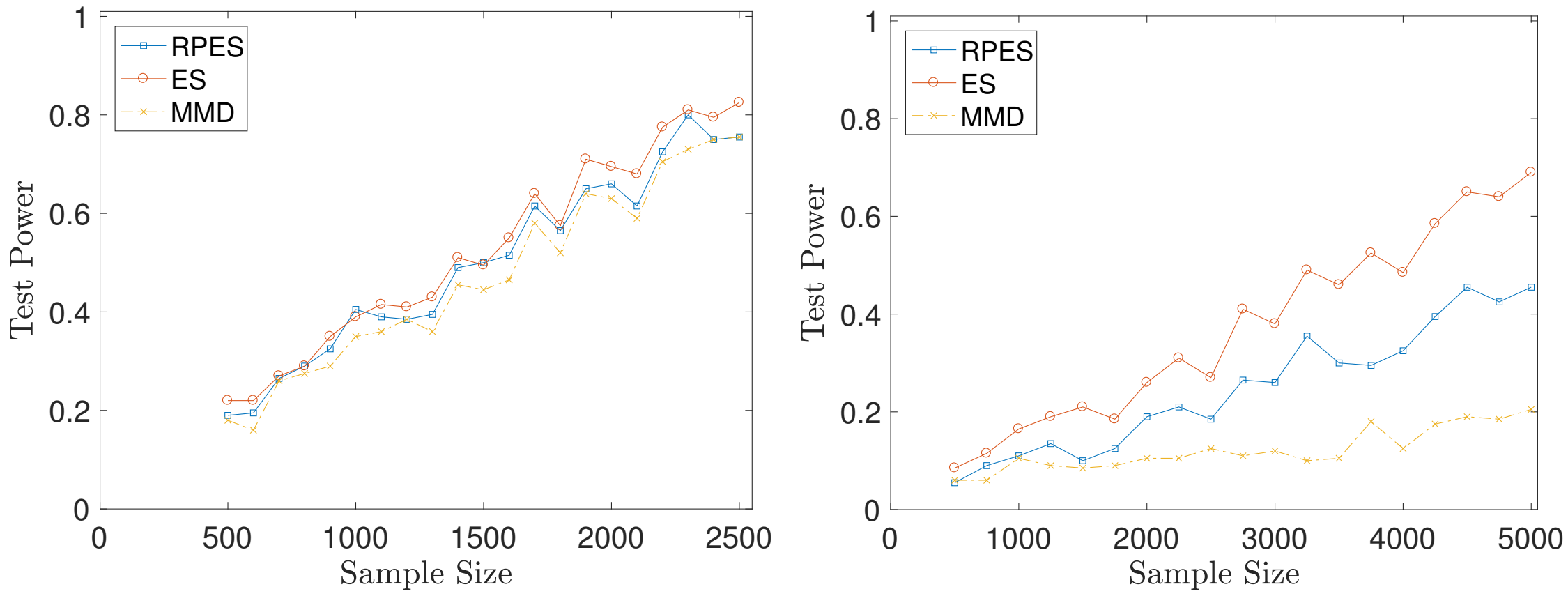

(a) $p = 5$ (b) $p = 50$

Figure 4: Test Power vs Sample Size in Example 5.3

Figure 4 plots the test power of each test against sample size in Example 5.3. In the low dimensional case, RPES, ES and MMD have similar performance. In higher dimensional case, RPES is less effective than ES since random projection may lose some efficiency when the mean of two distributions only differ in a single dimension. But, RPES still outperforms MMD by a significant margin.

In the next example, we will check how those tests perform when there is only a minor difference in degrees of freedom of two multivariate student t-distribution.

**Example 5.4.** *We generate random vector* $X \sim t_{\nu_1}^{(50)}$, $Y \sim t_{\nu_2}^{(50)}$, *where each entry of* $X$ *follows t-distribution with degree of freedom,* $X_i \sim t_{\nu_1}$, *and* $Y_i \sim t_{\nu_2}$. *We let* $(\nu_1, \nu_2) = (4, 5)$ *and* $(\nu_1, \nu_2) = (7, 10)$, *respectively. In both cases, the sample size n is from 500 to 5000 with an increment 250.*

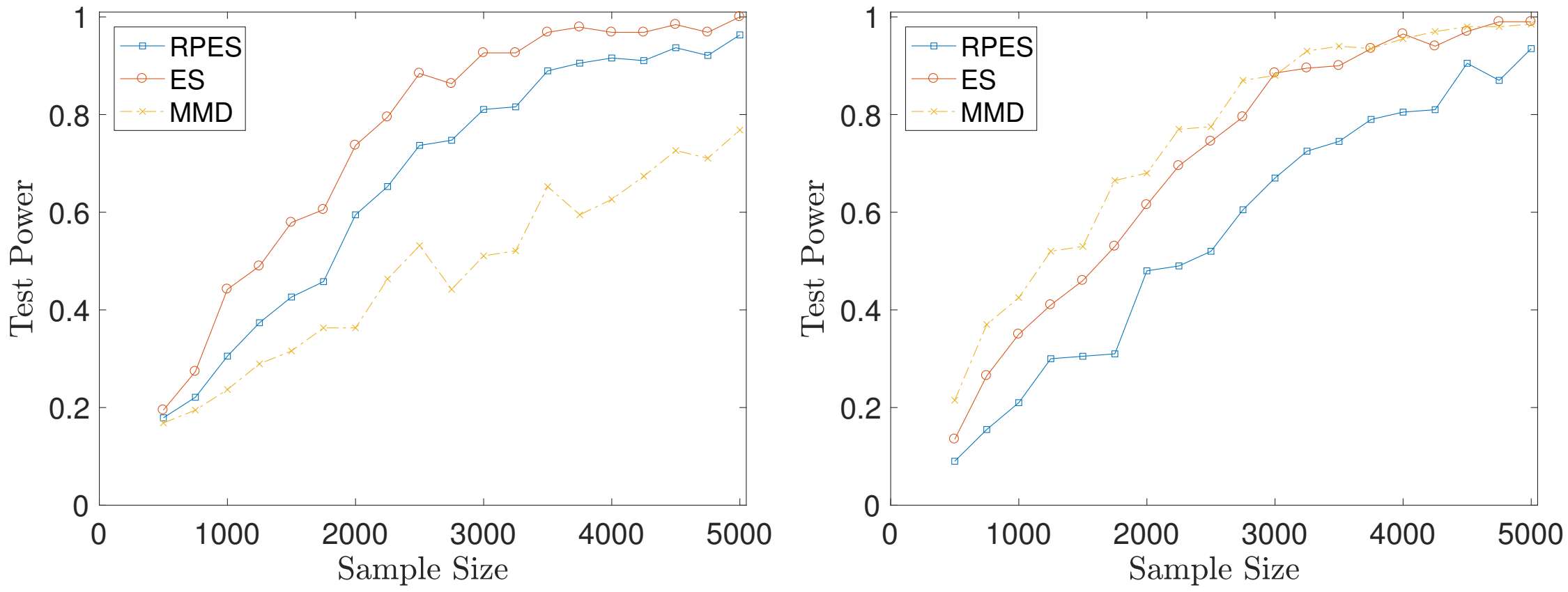

(a) $t_4^{(50)}$ v.s. $t_5^{(50)}$ (b) $t_7^{(50)}$ v.s. $t_{10}^{(50)}$

Figure 5: Test Power vs Sample Size in Example 5.4

Figure 5 plots the test power of each test against sample size in Example 5.4. In the first case, both RPES and ES outperforms MMD. In the second case, ES and MMD achieve similar performance while RPES underperforms slightly.

In the last example of this section, we will compare the performance of those tests in uniform distributions.

**Example 5.5.** *We generate random vector in the following two scenarios: (1) $X \sim$ Unif$(0,1)^{(5)}$, which means each entry of $X$ is drawn independently from Unif$(0,1)$, and $Y \sim$ Unif$(0,0.98)^{(5)}$; (2) $X \sim$ Unif$(0,1)^{(50)}$, and $Y \sim$ Unif$(0,0.99)^{(50)}$. In both cases, the sample size n is from 500 to 5000 with an increment 250.*

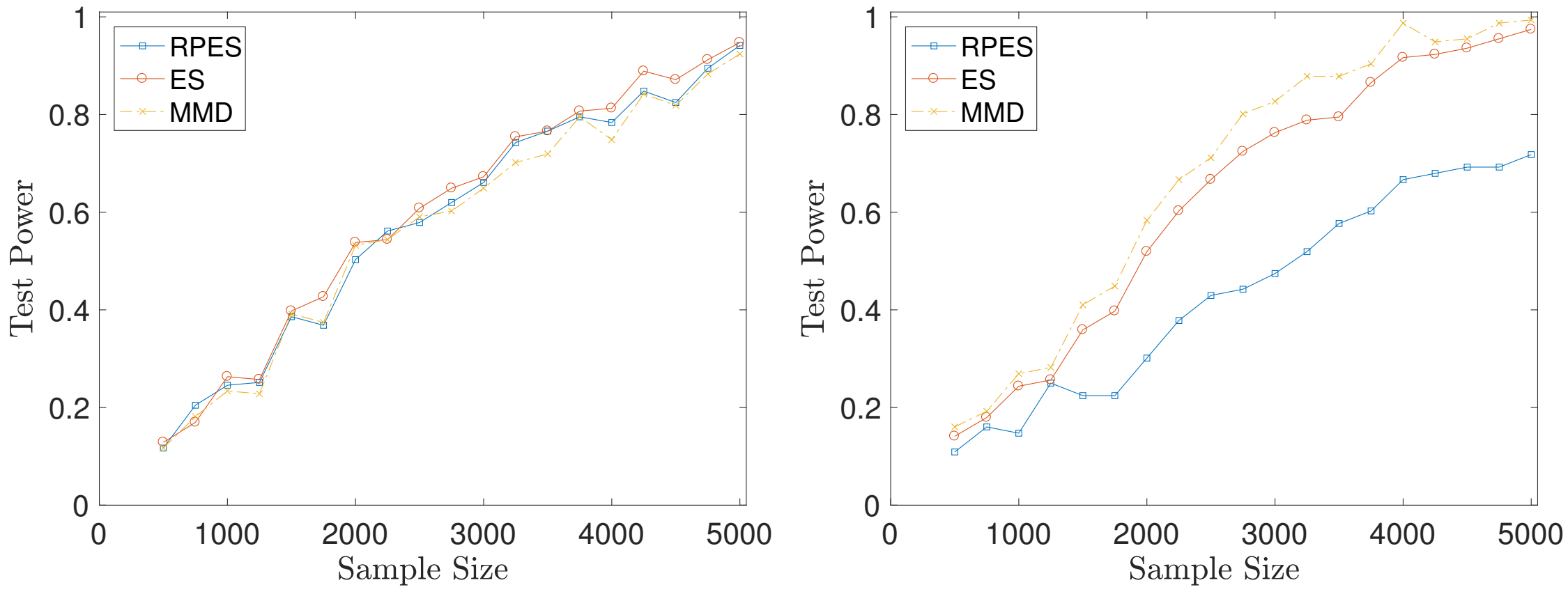

(a) $\mathrm{Unif}(0,1)^{(5)}$ v.s. $\mathrm{Unif}(0,0.98)^{(5)}$ (b) $\mathrm{Unif}(0,1)^{(50)}$ v.s. $\mathrm{Unif}(0,0.99)^{(50)}$

Figure 6: Test Power vs Sample Size in Example 5.5

Figure 6 plots the test power of each test against sample size in Example 5.5. Similar with the result of Example 5.3, the performance of RPES, ES and MMD are quite close in the lower dimensional case. In higher dimensional case, RPES and MMD are also very close in performance while RPES underperforms the aforementioned two methods.

The experiments results in this part show that ES performs best in nearly all the cases. Although RPES tends to be slightly less effective than ES when the data dimension is high and sample size is relatively small, their performances are quite close when the dimension is moderate or the size is sufficiently large.

# 6 Discussion

There are plenty existing work on graph-based two-sample tests. For instance, [5], [6] propose a graph-based two-sample test based on minimum spanning tree for multivariate data and categorical data, respectively. However, like aforementioned graph-based methods, they still suffer from the high computational complexity — $O(N^2 \log N)$ with Kruskal's algorithm. It is worth noting that [2] introduce a general notion of graph-based two-sample tests, and provide a unified framework for analyzing their asymptotic properties.

The kernel two-sample test statistic proposed by [8] has a very similar form with energy statistics. Though the Euclidean distance $f(x, y) = |x - y|$ is not a positive definite kernel, [19] show that distance-based methods and kernel-based methods might be unified under the same framework.

A possible application of the proposed two-sample tests is change-point detection. [18] develop a change-point detection method based on the minimum non-bipartite matching, which could be regared as an extension of [17]. So, it might be of interest to extend energy distance based method for change-detection problems.

The technique of random projection could be beneficial in reducing the computational complexity without significant compromise in statistical efficiency. [9] propose an computationally and statistically efficient test of independence with the random projection and distance covariance, which, together with this paper, reveals the potential of random projection in all distance-based methods.

Another interesting application of energy distance is distribution representation. [16] introduce a new way to compact a continuous probability distribution into a set of representative points called support points, which are obtained by minimizing the energy distance.

# 7 Conclusion

This paper makes three major contributions. First, we develop an efficient algorithm based on sorting and rearrangement to compute energy statistics for univariate random variables. Second, we propose an efficient scheme for computing the energy statistics of multivariate random variables with random projections and univariate fast algorithm. Third, we carry out a two-sample test based on the efficient algorithms and derive its asymptotic properties.

The theoretical analysis shows that the proposed test has nearly the same asymptotic efficiency (in terms of asymptotic variance) with the energy statistics. Numerical examples validate the theoretical results in computational and statistical efficiency.

# A Algorithms

We present all numerical algorithms of this paper here.

- Algorithm 1 summarizes how to compute the energy statistics of univariate random variables in $O(N \log N)$ time.
- Algorithm 2 describes how to approximate the energy statistics of random variables of any dimension in $O(KN \log N)$ time.
- Algorithm 3 describes a two-sample test that applies permutations to determine the threshold.
- Algorithm 4 describes a two-sample test using approximation of asymptotic distribution to determine the threshold.

**Algorithm 1:** A Fast Algorithm for Energy Statistics of Univariate Random Variables: $\mathcal{E}_{n,m}(X,Y)$

**Data**: Observations $X_1, \ldots, X_n \in \mathbb{R}$, $Y_1, \ldots, Y_m \in \mathbb{R}$;
**Result**: Energy Statistics $\mathcal{E}_{n,m}(X,Y)$

Sort $X_1, \ldots, X_n$ and $Y_1, \ldots, Y_m$. Let $X_{(1)} \leq X_{(2)} \leq \cdots \leq X_{(n)}$ and $Y_{(1)} \leq Y_{(2)} \leq \cdots \leq Y_{(m)}$ denote the order statistics.
Compute $E_2 = \frac{1}{n(n-1)} \sum_{i=1}^{n-1} i(n-i) \left|X_{(i+1)} - X_{(i)}\right|$ and $E_3 = \frac{1}{m(m-1)} \sum_{i=1}^{m-1} i(m-i) \left|Y_{(i+1)} - Y_{(i)}\right|$.
Merge two ordered series $X_{(i)}$'s and $Y_{(j)}$'s into a single ordered series $Z_{(1)} \leq \cdots \leq Z_{(n+m)}$. Let $I_i$ record the size of the subset of $Z_{(1)}$ through $Z_{(i)}$ that are from $X_{(i)}$'s.
Compute $E_1 = \frac{2}{nm} \sum_{i=1}^{n+m-1} \left[I_i(m-i+I_i) + (i-I_i)(n-I_i)\right] \left|Z_{(i+1)} - Z_{(i)}\right|$.
Return $\mathcal{E}_{n,m}(X,Y) = E_1 - E_2 - E_3$.

**Algorithm 2:** A Fast Algorithm for Energy Statistics of Multivariate Random Variables: $\overline{\mathcal{E}}_{m,n}$

**Data**: Observations $X_1, \ldots, X_n \in \mathbb{R}^p$, $Y_1, \ldots, Y_m \in \mathbb{R}^p$; Number of Random Projections $K$
**Result**: Average Randomly Projected Energy Statistics $\overline{\mathcal{E}}_{m,n}$

**for** *k = 1,…, K* **do**
  Randomly generate $u_k$ from Uniform($\mathcal{S}^{p-1}$);
  Compute the projection of $X_i$'s on $u_k$: $u_k^T X = (u_k^T X_1, \ldots, u_k^T X_n)$;
  Compute the projection of $Y_j$'s on $u_k$: $u_k^T Y = (u_k^T Y_1, \ldots, u_k^T Y_m)$;
  Compute the energy statistics of $u_k^T X$ and $u_k^T Y$ with Algorithm 1: $\mathcal{E}_{n,m}^{(k)} = C_p \mathcal{E}_{n,m}(u_k^T X, u_k^T Y)$;
**end**
Return $\overline{\mathcal{E}}_{m,n} = \frac{1}{K} \sum_{k=1}^{K} \mathcal{E}_{n,m}^{(k)}$.

**Algorithm 3:** Two-Sample Test Based on Permutations

**Data**: Observations $X_1, \ldots, X_n \in \mathbb{R}^p$, $Y_1, \ldots, Y_m \in \mathbb{R}^p$; Number of Random Projections $K$; Significance Level $\alpha_s$; Number of Permutations $L$

**Result**: Accept or Reject the Null Hypotheses $\mathcal{H}_0$: $X$ and $Y$ have the same distribution

Compute $\overline{\mathcal{E}}_{m,n}$ with Algorithm 2;

**for** $l$ = 1,..., $L$ **do**

  Generate a random permutation of the observations: $(X^{\star,l}, Y^{\star,l})$;

  Use Algorithm 2 to compute $D^{(l)} = \overline{\mathcal{E}}_{m,n}(X^{\star,l}, Y^{\star,l})$ with permutated observations;

**end**

Reject $\mathcal{H}_0$ if and only if $\frac{1+\sum_{l=1}^{L} I(\overline{\mathcal{E}}_{n,m} > D^{(l)})}{1+L} > \alpha_s$; otherwise, accept it.

**Algorithm 4:** Two-Sample Test Based on Approximated Asymptotic Distribution

**Data**: Observations $X_1, \ldots, X_n \in \mathbb{R}^p$, $Y_1, \ldots, Y_m \in \mathbb{R}^p$, $Z = (X_1, \ldots, X_n, Y_1, \ldots, Y_m)$; Number of Random Projections $K$; Significance Level $\alpha_s$

**Result**: Accept or Reject the Null Hypotheses $\mathcal{H}_0$: $X$ and $Y$ have the same distribution

**for** $k$ = 1,..., $K$ **do**

  Randomly generate $u_k$ from Uniform($\mathcal{S}^{p-1}$);

  Use Algorithm 1 to Compute:

  $\mathcal{E}_{n,m}^{(k)} = C_p \mathcal{E}_{n,m}(u_k^T X, u_k^T Y)$

  $S_{1;n,m}^{(k)} = C_p \binom{n+m}{2}^{-1} \sum_{i<j}^{n} |u^T(Z_i - Z_j)|$;

  Use the fast algorithm for distance covariance in [10] to compute:

  $S_{2;n,m}^{(k)} = C_p^2 \mathrm{SDC}(u_k^T Z, u_k^T Z)$;

  Randomly generate $v_k$ from Uniform($\mathcal{S}^{p-1}$);

  Use the fast algorithm for distance covariance in [10] to compute:

  $S_{3;n,m}^{(k)} = C_p^2 \mathrm{SDC}(u_k^T Z, v_k^T Z)$;

**end**

$\overline{\mathcal{E}}_{n,m} = \frac{1}{K}\sum_{k=1}^{K} \mathcal{E}_{n,m}^{(k)}$;  $\overline{S}_{1;n,m} = \frac{1}{K}\sum_{k=1}^{K} S_{1;n,m}^{(k)}$;

$\overline{S}_{2;n,m} = \frac{1}{K}\sum_{k=1}^{K} S_{2;n,m}^{(k)}$;  $\overline{S}_{3;n,m} = \frac{1}{K}\sum_{k=1}^{K} S_{3;n,m}^{(k)}$;

$\hat{\alpha} = \frac{1}{2} \frac{\overline{S}_{1;n,m}^2}{\frac{1}{K}\overline{S}_{2;n,m} + \frac{K-1}{K}\overline{S}_{3;n,m}}$;  $\hat{\beta} = \frac{1}{2} \frac{\overline{S}_{1;n,m}}{\frac{1}{K}\overline{S}_{2;n,m} + \frac{K-1}{K}\overline{S}_{3;n,m}}$;

Reject null hypothses $\mathcal{H}_0$ if and only if $(n+m)\overline{\mathcal{E}}_{n,m} + \overline{S}_{1;n,m} > \mathrm{Gamma}(1-\alpha_s; \hat{\alpha}, \hat{\beta})$; otherwise, accept it.

# B Proofs

We present all the proofs of this paper here.

## B.1 Proof of Theorem 3.1

*Proof.* The detailed explanations and corresponding complexity analysis of the fast algorithm in Section 3.1 is as follows.

(1) Sort $X_i$'s and $Y_j$'s, so that we have order statistics $X_{(1)} \le X_{(2)} \le \cdots \le X_{(n)}$ and $Y_{(1)} \le Y_{(2)} \le \cdots \le Y_{(m)}$. By adopting the merge sort [12, 11], the average computational complexity in this step is

$O(\max(n,m)\log\max(n,m))$. In addition, it is easy to verify the following:

$$\begin{aligned}\mathcal{E} := &\frac{2}{nm}\sum_{i=1}^{n}\sum_{j=1}^{m}|X_{(i)}-Y_{(j)}| - \frac{1}{n(n-1)}\sum_{i,j=1,i\neq j}^{n}|X_{(i)}-X_{(j)}| - \frac{1}{m(m-1)}\sum_{i,j=1,i\neq j}^{m}|Y_{(i)}-Y_{(j)}| \\ = &\frac{2}{nm}\sum_{i=1}^{n}\sum_{j=1}^{m}|X_{(i)}-Y_{(j)}| - \frac{2}{n(n-1)}\sum_{i<j}^{n}|X_{(i)}-X_{(j)}| - \frac{2}{m(m-1)}\sum_{i<j}^{m}|Y_{(i)}-Y_{(j)}|\end{aligned}$$

That is, we can compute $\mathcal{E}$ through merely the order statistics. The rest of algorithmic description will be based on the above formula.

(2) We can verify the following:

$$\frac{2}{n(n-1)}\sum_{i<j}^{n}|X_{(i)}-X_{(j)}| = \frac{2}{n(n-1)}\sum_{i=1}^{n-1} i(n-i)\left|X_{(i+1)}-X_{(i)}\right|.$$

Given order statistics $X_{(i)}$'s, the computational complexity of implementing the above is $O(n)$.

(3) Essentially identical to the previous item, one can verify the following:

$$\frac{2}{m(m-1)}\sum_{i<j}^{m}|Y_{(i)}-Y_{(j)}| = \frac{2}{m(m-1)}\sum_{i=1}^{m-1} i(m-i)\left|Y_{(i+1)}-Y_{(i)}\right|.$$

Given order statistics $Y_{(i)}$'s, the computational complexity of implementing the above is $O(m)$.

(4) For the first term in $\mathcal{E}$, one can computer it in two sub-steps as below.

(a) One can merge two ordered series $X_{(1)} \le X_{(2)} \le \cdots \le X_{(n)}$ and $Y_{(1)} \le Y_{(2)} \le \cdots \le Y_{(m)}$ into a single ordered series $Z_{(1)} \le Z_{(2)} \le \cdots \le Z_{(n+m)}$, where each $Z_{(k)}$ is either from $X_{(i)}$'s or from $Y_{(j)}$'s. At the same time, one can generate a sequence $I_i, i = 1, 2, \ldots, n+m$, where $I_i$ records the size of the subset of $Z_{(1)}$ through $Z_{(i)}$ that are from $X_{(i)}$'s. It is evident to show that quantity $i - I_i$ is the size of the subset of $Z_{(1)}$ through $Z_{(i)}$ that are from $Y_{(j)}$'s.
Note the computational complexity in this step is $O(n+m)$.

(b) Given the above preparation, we can verify the following:

$$\frac{2}{nm}\sum_{i=1}^{n}\sum_{j=1}^{m}\left|X_{(i)}-Y_{(j)}\right| = \frac{2}{nm}\sum_{i=1}^{n+m-1}\left[I_i(m-i+I_i)+(i-I_i)(n-I_i)\right]\left|Z_{(i+1)}-Z_{(i)}\right|.$$

Note that the term $I_i(m-i+I_i)+(i-I_i)(n-I_i)$ on the right hand side is equal to the number of times the length $|Z_{(i+1)}-Z_{(i)}|$ has been counted in the double summation on the left hand side. Through this, we can establish the equality. The computational complexity of implementing the above is $O(n+m)$.

From all the above, we show that the complexity of computing $\mathcal{E}$ is dominated by the sorting step, thus the average total complexity is $O(N\log N)$, where $N = n+m$. □

### B.2 Proof of Lemma 4.1

*Proof.* The proof is straightforward. First, by Proposition 2.2, we know that

random vector $X \in \mathbb{R}^p$ has the same distribution with random vector $Y \in \mathbb{R}^p$

if and only if

$$\Phi_X = \Phi_Y, \text{ almost everywhere,}$$

where $\Phi_X$ and $\Phi_Y$ are the characteristic functions of $X$ and $Y$, respectively. That becomes

$$\mathbb{E}\left[e^{iX^T t}\right] = \mathbb{E}\left[e^{iY^T t}\right], \forall t \in \mathbb{R}^p.$$

By variable change $t = ut'$, where $u \in \mathcal{S}^{p-1}$ and $t' \in [0, \infty)$, we have

$$\mathbb{E}\left[e^{iu^T X t'}\right] = \mathbb{E}\left[e^{iu^T Y t'}\right], \forall u \in \mathcal{S}^{p-1} \text{ and } t' \in [0, \infty),$$

or equivalently,

$$\Phi_{u^T X} = \Phi_{u^T Y}, \forall u \in \mathcal{S}^{p-1}.$$

By Proposition 2.2, we know that

$$\Phi_{u^T X} = \Phi_{u^T Y}, \forall u \in \mathcal{S}^{p-1},$$

is equivalent with

$$\mathcal{E}(u^T X, u^T Y) = 0, \forall u \in \mathcal{S}^{p-1}.$$

□

### B.3 Proof of Lemma 4.2

First, let us state a result from [9], which shows relationship between the norm of random projections and the norm of original vector.

**Lemma B.1.** *[9, Lemma B.1] Suppose $v$ is a fixed unit vector in $\mathbb{R}^p$ and $u \in \mathcal{S}^{p-1}$. Let $\mu$ be the uniform probability measure on $\mathcal{S}^{p-1}$. We have*

$$C_p \int_{\mathcal{S}^{p-1}} |u^T v| d\mu(u) = C_p \mathbb{E}_u[|u^T v|] = 1,$$

*where constant $C_p$ has been mentioned at the end of Section 1.*

Equipped with above lemma, we can prove Lemma 4.2 as follows.

*Proof.* By Lemma B.1, we have

$$C_p \mathbb{E}_u \left[\left| u^T \frac{(X-Y)}{|X-Y|} \right|\right] = 1, \text{ thus, } |X-Y| = C_p \mathbb{E}_u \left[\left|u^T(X-Y)\right|\right].$$

Therefore, the energy distance could be written as

$$\begin{aligned}
\mathcal{E}(X,Y) &= 2\mathbb{E}[|X-Y|] - \mathbb{E}[|X-X'|] - \mathbb{E}[|Y-Y'|] \\
&= 2\mathbb{E}_{X,Y}[C_p \mathbb{E}_u[|u^T(X-Y)]] - \mathbb{E}_{X,X'}[C_p \mathbb{E}_u[|u^T(X-X')]] - \mathbb{E}_{Y,Y'}[C_p \mathbb{E}_u[|u^T(Y-Y')]] \\
&= C_p \mathbb{E}_u \left[2\mathbb{E}_{X,Y}[|u^T(X-Y)] - \mathbb{E}_{X,X'}[|u^T(X-X')] - \mathbb{E}_{Y,Y'}[|u^T(Y-Y')]\right] \\
&= C_p \mathbb{E}_u[\mathcal{E}(u^T X, u^T Y)] = C_p \int_{\mathcal{S}^{p-1}} \mathcal{E}(u^T X, u^T Y) d\mu(u),
\end{aligned}$$

where $u$ is a uniformly distributed random variable on $\mathcal{S}^{p-1}$, the second equality is by Lemma B.1, the third equality is by exchanging the order of expectation, and the fourth equality is by the definition of energy distance.

We can reach a similar result for energy statistics simply by replacing $\mathbb{E}_{X,Y}[\cdot]$, $\mathbb{E}_{X,X'}[\cdot]$ and $\mathbb{E}_{Y,Y'}[\cdot]$ with summation. The rest reasoning is almost the same with above reasoning for energy distance. □

## B.4 Proof of Theorem 4.7

First, let us introduce a lemma that will be used in later proof.

**Lemma B.2.** *[9, Lemma 4.13] If $\mathbb{E}[|X|^2] < \infty$, we have that kernel*

$$\mathbf{k}(X_1, X_2) = \mathbb{E}_X[|X_1 - X|] + \mathbb{E}_X[|X_2 - X|] - |X_1 - X_2| - \mathbb{E}_{X,X'}[|X - X'|]$$

*is a positive definite kernel. As a result, if $X$ and $Y$ have the same distirbution, $h_{20}(\cdot,\cdot)$, $h_{02}(\cdot,\cdot)$ and $-h_{11}(\cdot,\cdot)$ in Lemma 4.6 are all positive definite kernels. Also, there exist functions $\phi_1(\cdot), \phi_2(\cdot), \ldots$ such that*

$$\mathbf{k}(X_1, X_2) = \sum_{i=1}^{\infty} \lambda_i \phi_i(X_1)\phi_i(X_2),$$

*where $\lambda_1 \geq \lambda_2 \geq \ldots \geq 0$, $\mathbb{E}[\phi_i(X)] = 0$, $\mathbb{E}[\phi_i(X)^2] = 1$ and $\mathbb{E}[\phi_i(X)\phi_j(X)] = 0$, $i = 1, 2, \ldots, \infty$, $i \neq j$.*

Now, let us prove Theorem 4.7.

*Proof.* By Lemma 4.6 and [14, Section 2.2, Theorem 3], we have

$$\mathcal{E}_{n,m} = \binom{n}{2}^{-1} \sum_{i_1 < i_2} h_{20}(X_{i_1}, X_{i_2}) + 4(nm)^{-1} \sum_{i=1}^{n}\sum_{j=1}^{m} h_{11}(X_i, Y_j) + \binom{m}{2}^{-1} \sum_{j_1 < j_2} h_{02}(Y_{j_1}, Y_{j_2}) + \mathcal{R}_{n,m},$$

where $\mathcal{R}_{n,m}$ is the residual with $N\mathcal{R}_{n,m} \xrightarrow{P} 0$. By Lemma B.2, we know that

$$h_{20}(X_{i_1}, X_{i_2}) = \sum_{l=1}^{\infty} \lambda_l \phi_l(X_{i_1})\phi_l(X_{i_2}), \qquad h_{02}(Y_{j_1}, Y_{j_2}) = \sum_{l=1}^{\infty} \lambda_l \phi_l(Y_{j_1})\phi_l(Y_{j_2}),$$

and

$$h_{11}(X_i, Y_j) = -\frac{1}{2}\sum_{l=1}^{\infty} \lambda_l \phi_l(X_i)\phi_l(Y_j).$$

Therefore, we have

$$\begin{aligned}
\mathcal{E}_{n,m} &= \sum_{l=1}^{\infty} \lambda_l \left[ \left( \frac{1}{n}\sum_{i=1}^{n} \phi_l(X_i) - \frac{1}{m}\sum_{j=1}^{m} \phi_l(Y_j) \right)^2 - \frac{1}{n^2}\sum_{i=1}^{n} \phi_l(X_i)^2 - \frac{1}{m^2}\sum_{j=1}^{m} \phi_l(Y_j)^2 \right] \\
&\quad + \mathcal{R}_{n,m} + \left( \frac{2}{n(n-1)} - \frac{2}{n^2} \right) \sum_{i_1 < i_2} h_{20}(X_{i_1}, X_{i_2}) + \left( \frac{2}{m(m-1)} - \frac{2}{m^2} \right) \sum_{j_1 < j_2} h_{02}(Y_{j_1}, Y_{j_2}) \\
&= \frac{1}{N}\sum_{l=1}^{\infty} \lambda_l \left[ \left( \sqrt{N/n}\frac{1}{\sqrt{n}}\sum_{i=1}^{n} \phi_l(X_i) - \sqrt{N/m}\frac{1}{\sqrt{m}}\sum_{j=1}^{m} \phi_l(Y_j) \right)^2 \right. \\
&\quad \left. - \frac{N}{n^2}\sum_{i=1}^{n} \phi_l(X_i)^2 - \frac{N}{m^2}\sum_{j=1}^{m} \phi_l(Y_j)^2 \right] + \tilde{\mathcal{R}}_{n,m},
\end{aligned}$$

where

$$\tilde{\mathcal{R}}_{n,m} = \mathcal{R}_{n,m} + \frac{2}{n^2(n-1)} \sum_{i_1<i_2} h_{20}(X_{i_1}, X_{i_2}) + \frac{2}{m^2(m-1)} \sum_{j_1<j_2} h_{02}(Y_{j_1}, Y_{j_2}).$$

It is worth noting that $N\tilde{\mathcal{R}}_{n,m} \xrightarrow{P} 0$. Therefore, as $N \to \infty$, we have

$$N\mathcal{E}_{n,m} \xrightarrow{D} \sum_{l=1}^{\infty} \lambda_l[(\sqrt{1/\eta}Z_{l,1} - \sqrt{1/(1-\eta)}Z_{l,2})^2 - \frac{1}{\eta} - \frac{1}{1-\eta}] = \sum_{l=1}^{\infty} \frac{\lambda_l}{\eta(1-\eta)}(Z_l^2 - 1),$$

where $Z_{l,1}, Z_{l,2}, l = 1, 2, \ldots$ are all independent standard normal random variables and $Z_l = \sqrt{1-\eta}Z_{l,1} + \sqrt{\eta}Z_{l,2}$. It is worth noting that

$$\sum_{l=1}^{\infty} \lambda_l = \mathbb{E}[h_{20}(X, X)] = \mathbb{E}[|X - X'|].$$

Similarly, we know that

$$\begin{aligned}
\sum_{l=1}^{\infty} \lambda_l^2 &= \mathbb{E}_{X_1,X_2}\left[\sum_{l=1}^{\infty} \lambda_l \phi_l(X_1)\phi_l(X_2)\right]^2 \\
&= \mathbb{E}_{X_1,X_2}\left[h_{20}(X_1, X_2)^2\right] \\
&= \mathbb{E}_{X_1,X_2}\left[\left(\mathbb{E}_X[|X_1 - X|] + \mathbb{E}_X[|X_2 - X|] - |X_1 - X_2| - \mathbb{E}_{X,X'}[|X - X'|]\right)^2\right] \\
&= DC(X, X),
\end{aligned}$$

where the last equality is by the definition of distance covariance. □

## B.5 Proof of Lemma 4.10

*Proof.* It is worth noting that when $X$ and $Y$ have the same distribution, $u^T X$ and $u^T Y$ also should have the same distribution for any $u$, thus

$$\mathcal{E}(u^T X, u^T Y) = 0, \forall u \in \mathbb{R}^p,$$

which indicates that

$$\text{Var}_u[\mathcal{E}(u^T X, u^T Y)] = 0.$$

Moreover, we have

$$\overline{h}_{10}(X_1) = \frac{1}{K}\sum_{k=1}^{K} C_p h_{10}(u_k^T X_1).$$

By the definition of $h_{10}(\cdot)$, we know $h_{10}(u_k^T X_1) = 0$ when $X$ and $Y$ are identically distributed, which sugguests

$$\overline{h}_{10} = 0, \text{ and } \text{Var}[\overline{h}_{10}|U] = 0.$$

Similarly, we have

$$\overline{h}_{01} = 0, \text{ and } \text{Var}[\overline{h}_{01}|U] = 0.$$

Combining above results and Lemma 4.8, we have the formula of the variance of $\overline{\mathcal{E}}_{n,m}$ in this lemma. □

### B.6 Proof of Theorem 4.12

*Proof.* This proof is almost identical with the proof of Theorem 4.7. We can simply replace the notations like $h_{20}, h_{02}, h_{11}, \lambda_i, \phi_i(\cdot)$ with corresponding notations like $\overline{h}_{20}, \overline{h}_{02}, \overline{h}_{11}, \overline{\lambda}_i, \overline{\phi}_i(\cdot)$. The rest reasoning is the same.

For $\sum_{l=1}^{\infty} \overline{\lambda}_l$, it is easy to see that

$$\sum_{l=1}^{\infty} \overline{\lambda}_l = \mathbb{E}[\overline{\mathbf{k}}(X, X)] = \frac{C_p}{K} \sum_{k=1}^{K} \mathbb{E}_{X,X'}[|u_k^T(X - X')|].$$

For $\sum_{l=1}^{\infty} \overline{\lambda}_l^2$, we have

$$\begin{aligned}
\sum_{l=1}^{\infty} \overline{\lambda}_l^2 &= \mathbb{E}_{X_1,X_2}\left[\sum_{l=1}^{\infty} \overline{\lambda}_l \overline{\phi}_l(X_1)\overline{\phi}_l(X_2)\right]^2 = \mathbb{E}_{X_1,X_2}\left[C_p^2 \overline{\mathbf{k}}(u_k^T X_1, X_2)^2\right] \\
&= \mathbb{E}_{X_1,X_2}\left[\left(\frac{C_p^2}{K} \sum_{k=1}^{K} \mathbf{k}(u_k^T X_1, u_k^T X_2)\right)^2\right] \\
&= \frac{C_p^2}{K^2} \sum_{k,k'=1}^{K} \mathbb{E}_{X_1,X_2}\left[\mathbf{k}(u_k^T X_1, u_k^T X_2)\mathbf{k}(u_{k'}^T X_1, u_{k'}^T X_2)\right] \\
&= \frac{C_p^2}{K^2} \sum_{k,k'=1}^{K} DC(u_k^T X, u_{k'}^T X),
\end{aligned}$$

where the last equation is by the definition of distance covariance. □

### B.7 Proof of Proposition 4.13

*Proof.* When $X$ and $Y$ are identically distributed, we know

$$\mathbb{E}[|u_k^T(Z_i - Z_j)|] = \mathbb{E}_{X,X'}[|u_k^T(X - X')|],$$

which implies

$$\frac{C_p}{K} \sum_{k=1}^{K} \frac{1}{(n+m)(n+m-1)} \sum_{i \neq j}^{n+m} |u_k^T(Z_i - Z_j)|$$

is an unbiased estimator for $\frac{C_p}{K} \sum_{k=1}^{K} \mathbb{E}_{X,X'}[|u_k^T(X - X')|] = \sum_{l=1}^{\infty} \overline{\lambda}_l$.

We have

$$\begin{aligned}
\sum_{l=1}^{\infty} \overline{\lambda}_l^2 &= \frac{C_p^2}{K^2} \sum_{k,k'=1}^{K} \text{DC}(u_k^T X, u_{k'}^T X) \\
&= \frac{C_p^2}{K^2} \sum_{k=1}^{K} \text{DC}(u_k^T X, u_k^T X) + \frac{C_p^2}{K^2} \sum_{k \neq k'}^{K} \text{DC}(u_k^T X, u_{k'}^T X).
\end{aligned}$$

For $\text{DC}(u_k^T X, u_k^T X)$ in the first term , it is natural to estimate it with $\text{SDC}(u_k^T Z, u_k^T Z)$. It is worth noting that $u_k$ is independent of $u_{k'}$ for all $k' \neq k$. When the number of random projections $K$ is sufficient large, by the Law of Large Number, we have

$$\frac{C_p^2}{K^2}\sum_{k\neq k'}^{K}\text{DC}(u_k^TX, u_{k'}^TX) \xrightarrow{P} \frac{(K-1)C_p^2}{K^2}\sum_{k=1}^{K}\text{DC}(u_k^TX, v_k^TX).$$

We can estimate the quantity on the right-hand-side by simply estimating distance covariance with the sample version. Thus, we have

$$\frac{C_p^2}{K^2}\sum_{k=1}^{K}\text{SDC}(u_k^TZ, u_k^TZ) + \frac{(K-1)C_p^2}{K^2}\sum_{k=1}^{K}\text{SDC}(u_k^TZ, v_k^TZ) \to \sum_{l=1}^{\infty}\overline{\lambda}_l^2 \text{ as } N, K \to \infty.$$

□